\pdfoutput=1

\documentclass{aa}   

\usepackage{graphicx,natbib,url,twoopt}
\usepackage[varg]{txfonts}           
\usepackage{hyperref}   
\usepackage{url}            
\usepackage{pdfcomment}              
\usepackage{acronym} 
\usepackage{epstopdf}                
\usepackage{bm}

\hypersetup{
  colorlinks=true,   
  urlcolor=blue,     
  linkcolor=red,     
}

\bibpunct{(}{)}{;}{a}{}{,}    

\makeatletter
\newcommand{\bibnote}[2]{\global\@namedef{#1note}{#2}}
\newcommand{\biblink}[2]{\global\@namedef{#1link}{#2}}
\makeatother

\makeatletter
 \newcommandtwoopt{\citeads}[3][][]{%
   \nonstopmode
   \href{http://adsabs.harvard.edu/abs/#3}%
        {\def\hyper@linkstart##1##2{}%
         \let\hyper@linkend\@empty\citealp[#1][#2]{#3}}
   \biblink{#3}{\href{http://adsabs.harvard.edu/abs/#3}{ADS}}%
   \errorstopmode}            
 \newcommandtwoopt{\citepads}[3][][]{%
   \nonstopmode
   \href{http://adsabs.harvard.edu/abs/#3}%
        {\def\hyper@linkstart##1##2{}%
         \let\hyper@linkend\@empty\citep[#1][#2]{#3}}
   \biblink{#3}{\href{http://adsabs.harvard.edu/abs/#3}{ADS}}%
   \errorstopmode}            
 \newcommandtwoopt{\citetads}[3][][]{%
   \nonstopmode
   \href{http://adsabs.harvard.edu/abs/#3}%
        {\def\hyper@linkstart##1##2{}%
         \let\hyper@linkend\@empty\citet[#1][#2]{#3}}
   \biblink{#3}{\href{http://adsabs.harvard.edu/abs/#3}{ADS}}%
   \errorstopmode}            
 \newcommandtwoopt{\citeyearads}[3][][]{%
   \nonstopmode
   \href{http://adsabs.harvard.edu/abs/#3}%
        {\def\hyper@linkstart##1##2{}%
         \let\hyper@linkend\@empty\citeyear[#1][#2]{#3}}
   \biblink{#3}{\href{http://adsabs.harvard.edu/abs/#3}{ADS}}%
   \errorstopmode}            
\makeatother

\newacro{ADS}{Astrophysics Data System}
\newacro{NLTE}{non-local thermodynamic equilibrium}
\newacro{NASA}{National Aeronautics and Space Administration}

\begin{document}  

\title{Anisotropy of the galaxy cluster X-ray luminosity-temperature relation}

\author{Konstantinos Migkas \and Thomas H. Reiprich}
\institute{Argelander-Institut f{\"u}r Astronomie, Universit{\"a}t Bonn, Auf dem H{\"u}gel 71, 53121 Bonn, Germany \\ \email{kmigkas@astro.uni-bonn.de}
}

\date{Accepted date: 30/10/2017} 

\abstract{We introduce a new test to study the Cosmological Principle with galaxy clusters. Galaxy clusters exhibit a tight correlation between the luminosity and temperature of the X-ray-emitting intracluster medium. While the luminosity measurement depends on cosmological parameters through the luminosity distance, the temperature determination is cosmology-independent. We exploit this property to test the isotropy of the luminosity distance over the full extragalactic sky, through the normalization $a$ of the $L_X-T$ scaling relation and the cosmological parameters $\Omega_{\text{m}}$ and $H_0$. To this end, we use two almost independent galaxy cluster samples: the ASCA Cluster Catalog (ACC) and the XMM Cluster Survey (XCS-DR1). Interestingly enough, these two samples appear to have the same pattern for $a$ with respect to the Galactic longitude. More specifically, we identify one sky region within $l\sim (-15^{\circ},90^{\circ})$ (Group A) that shares very different best-fit values for the normalization of the $L_X-T$ relation for both ACC and XCS-DR1 samples. We use the Bootstrap and Jackknife methods to assess the statistical significance of these results. We find the deviation of Group A, compared to the rest of the sky in terms of $a$, to be $\sim 2.7\sigma$ for ACC and $\sim 3.1\sigma$ for XCS-DR1. This tension is not significantly relieved after excluding possible outliers and is not attributed to different redshift ($z$), temperature ($T$), or distributions of observable uncertainties. Moreover, a redshift conversion to the cosmic microwave background (CMB) frame does not have an important impact on our results. Using also the HIFLUGCS sample, we show that a possible excess of cool-core clusters in this region, is not able to explain the obtained deviations. Furthermore, we tested for a dependence of the results on supercluster environment, where the fraction of disturbed clusters might be enhanced, possibly affecting the $L_X-T$ relation. We indeed find a trend in the XCS-DR1 sample for supercluster members to be underluminous compared to field clusters. However, the fraction of supercluster members is similar in the different sky regions, so this cannot explain the observed differences, either. Constraining $\Omega_{\text{m}}$ and $H_0$ via the redshift evolution of $L_X-T$ and the luminosity distance via the flux-luminosity conversion, we obtain approximately the same deviation amplitudes as for $a$. It is interesting that the general observed behavior of $\Omega_{\text{m}}$ for the sky regions that coincide with the CMB dipole is similar to what was found with other cosmological probes such as supernovae Ia.
The reason for this behavior remains to be identified.}

\keywords{Cosmology: observations -- X-rays:galaxies:clusters -- galaxies: clusters: general -- methods: statistical}

\maketitle

\section{Introduction} \label{intro}


The Cosmological Principle (CP) is considered to be the foundation of modern cosmology, stating that the Universe must be isotropic and homogeneous on sufficiently large scales. It is robustly supported by various cosmological probes such as the cosmic microwave background observed by WMAP \citep{bennett} and Planck \citep{planck} satellites, the distribution of distant radio sources \citep{condon,blake} and the large scale distribution of galaxies \citep{marinoni,appleby,alonso,pandey}. 

However, several studies, using the magnitude-redshift relation of Type Ia Supernovae (SNIa), have reported mild-significance anisotropic signals of the Hubble expansion, mostly correlated with the cosmic microwave background (CMB) dipole \citep{schwarz,antoniou,mariano,appleby-john,bengaly15,javanm15,migkas}, even if the redshifts of the SNIa have been adjusted to the CMB rest frame prior to the analysis. However, such signals can be attributed to single SNIa acting as outliers, that can affect the cosmological parameters derived from smaller subsamples, as shown in \citet{migkas}.
 
Furthermore, a similar dipole anisotropy has been found in the X-ray background from previous studies \citep{shafer, plionis99}, that could again be attributed to the local motions of the Local Group.

Of course, the consistency of the pinpointed anisotropy signal in different SNIa samples and other independent probes must be further investigated. A systematic finding of a dipole anisotropy that coincides with the CMB dipole direction could indicate that the reason behind the latter is not exclusively due to the Doppler shift caused by our own bulk motion.

The CP must be valid not only for cosmological parameters, but for the properties of astrophysical objects as well. 
\citet{javanm17} found a significant hemisphere anisotropy in the galaxy morphological types aligned with the rotational axis of the Earth, arguing that is probably caused by a systematic bias in the classification of galaxy types.

Some of the most interesting objects to study and use in order to trace the behavior of the large-scale structure are galaxy clusters. They are the largest gravitationally bound systems in the universe, easily detected in the X-ray regime due to the large amounts of hot gas they contain ($\sim 10$\% of their total mass) in the intra-cluster medium (ICM). One of the most crucial properties of galaxy clusters are their scaling relations, correlating important physical quantities such as luminosity, temperature, and mass with each other. 
\citet{kaiser}, based on the self-similar model, provided a theoretical prediction for these scaling laws. 

Specifically, the relation between the X-ray luminosity and the temperature of the ICM gas is given by $L_X \ E(z)^{-1}\propto T^2$, where $E(z)=\sqrt{\Omega_{\text{m}} (1+z)^3+(1-\Omega_{\text{m}}-\Omega_{\Lambda})(1+z)^2+\Omega_{\Lambda}}$ takes into account the redshift evolution of the relation \citep{giodini}. This is derived under the assumption that gravitational energy is the only source of energy transferred to the ICM. The $L_X-T$ relation has been well-investigated \citep{edge,markev,vikhl02,pacaud07,pratt,eckmiller,mittal,hilton,maughan,takey,connor,bharad,lovisari} . The observed slope of the power-law is systematically steeper than the predicted one, indicating the existence of different energy sources contributing to the ICM. Such sources are active galactic nuclei (AGN) feedback, supernovae-driven winds, and so on.

While the $L_X-T$ slope heavily depends on the physical processes that heat the gas in the ICM, differences in the normalization of the relation can potentially reflect any differences that might exist in the cosmological parameters for different directions on the sky. 
This is due to the fact that $H_0, \ \Omega_{\text{m}}, \ \Omega_{\Lambda}$ , and so on, enter through the conversion of the observed X-ray flux to the X-ray luminosity, as well as from the $E(z)$ factor. Therefore, for fixed values of redshift and temperature, a higher value for $\Omega_{\text{m}}$ towards a sky region would lead to lower luminosity distances, thus to lower X-ray luminosities, and eventually to lower normalization values.
At the same time, the temperature determination is cosmology-independent, something that motivates the use of the $L_X-T$ relation for anisotropy studies. Assuming fixed values for the cosmological parameters, one could precisely determine the normalization and slope of $L_X-T$ or vice versa. 
All these make galaxy clusters excellent tools for constraining the cosmological parameters and studying their underlying physics.

Previous studies have used the kinematic Sunyaev-Zeldovic effect of  galaxy clusters to trace the large scale peculiar motions up to  $\sim 600\ h^{-1}$ Mpc, reporting challenging results for the $\Lambda$CDM model \citep{kashl08,kashl10,kash11,atrio-bar}. However, the significance of these results has been challenged due to the controversial validity of the applied method \citep{keisler,osborne,planck14}. Moreover, the X-ray flux-weighted  method
has been used for galaxy cluster luminosity functions \citep{plionis98,kocevski}, finding consistent results with the concordance cosmology.

In addition, \citet{bengaly17-clusters} used the Planck measurements of the Sunyaev-Zeldovic effect \citep{ade-c,ade-b} to probe the angular distribution of clusters in antipodal patches of the sky, finding fully consistent results with the statistical isotropy assumption.

In this study, we introduce a new method to test the validity of the CP, namely the isotropy of the $L_X-T$ scaling relation of galaxy clusters. To this end, we use two almost independent galaxy cluster samples. The first is contained in the Advanced Satellite for Cosmology and Astrophysics (ASCA) \citep{tanaka} catalogue and was compiled by D. Horner \citep{horner} under the name ASCA Cluster Catalog (ACC); the second is the first data release from the XMM-Newton Cluster Survey (XCS-DR1) \citep{mehrt}. Throughout this paper, we correct $L_X$ values, as given by the data samples, to a $\Lambda$CDM cosmology with $H_0=70\  \text{km}\  \text{s}^{-1} \text{Mpc}^{-1}$, $\Omega_{\text{m}}=0.28$ and $\Omega_{\Lambda}=0.72$. Finally, $\log{x}$ is used as $\log_{10}{x}$. 

\section{Data samples}

\subsection{ACC} \label{acc_sample}

We use all the 272 galaxy clusters and groups contained in ACC, for which we have information for their right ascension (R.A.), declination (DEC), redshift ($z$), bolometric X-ray luminosity $(L_X)$, correction factor $l_{\text{vir}}$ to apply to $L_X$ to obtain the luminosity within the virial radius $r_{200}$ (the radius inside which the mean density of the cluster is 200 times greater than the critical density of the Universe, e.g., the radius where we consider the virialized halo to extend) and X-ray temperature $(T)$ with its 90\%-confidence levels.
The confidence levels for $L_X$ are not given. The extraction region of the spectra and $L_X$, was chosen such that the radial profile of cluster counts was at least $5\sigma$ greater than the background signal; for the vast majority of the objects, this corresponds to a radius of $\ \sim 0.8\ r_{200}-1.05\ r_{200}$. In order to be consistent for all the clusters of the sample, we use the bolometric $L_X$ emitted from within $r_{200}$.  Moreover, while the absorbed $L_X$ is given for all the 272 objects, only 230 clusters are also given with their $L_X$ values corrected for the neutral hydrogen column density absorption. Therefore, using \textsc{XSPEC} \citep{xspec} we correct the $L_X$ values ourselves for the remaining 42 galaxy clusters and groups.

For these 272 objects, the median relative temperature uncertainty is $\left(\dfrac{\sigma_T}{T}\right)=6.3\%$, demonstrating the spectroscopic precision of the observations.

\begin{figure}[hbtp]
                \resizebox{\hsize}{!}{\includegraphics{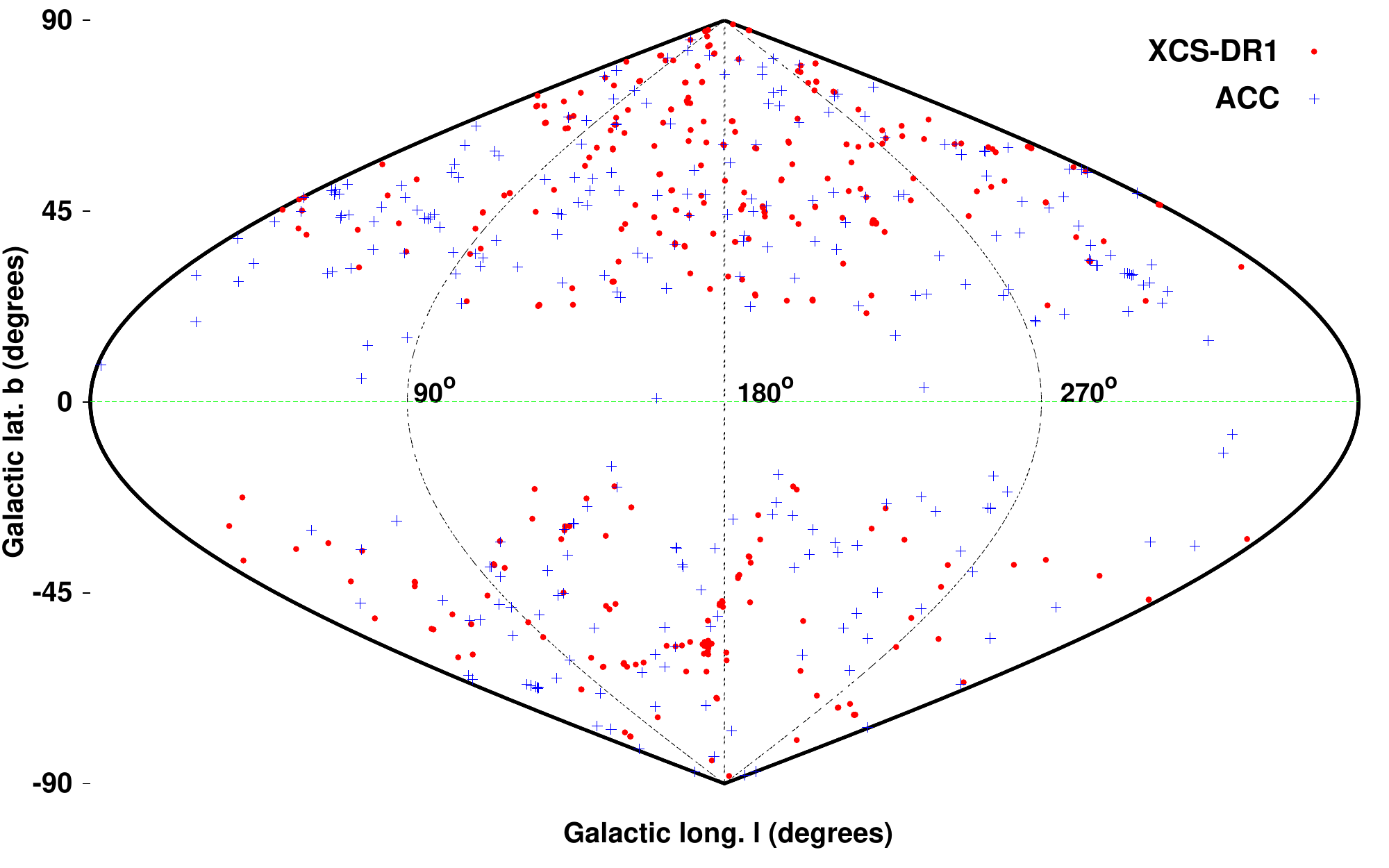}}
        \caption{Positions of the galaxy clusters contained in ACC (blue) and XCS-DR1 (red) at the Galactic sky map.}
        \label{fig1}
\end{figure}

A specific selection function has not been applied to the ASCA clusters, since the final sample is an archival one composed of observations obtained for the needs of different projects at different times and not by one full-sky survey. To this preliminary catalog, \citet{horner} applied a homogeneous data reduction pipeline to obtain the final ACC sample.
The spatial distribution of the ACC clusters is displayed in Fig. \ref{fig1}. Finally, the cosmology used to derive $L_X$ was an EdS universe with $H_0=50\  \text{km}\  \text{s}^{-1}\ \text{Mpc}^{-1},\ \Omega_{\text{m}}=1,$ and $\Omega_{\Lambda}=0$, which we convert to our default cosmology.

\subsection{XCS-DR1}

This sample consists of 503 optically confirmed X-ray galaxy clusters, serendipitously  detected by the XMM-Newton telescope \citep{xmm} and drawn from publicly available data \citep{mehrt}. The galaxy clusters were homogeneously selected, covering the full sky (except for the Galactic latitudes $|b|\le 20^{\circ}$). Out of these 503 clusters, 356 are observed in X-rays for the first time and 255 are newly discovered.

We make use of the 364 clusters for which the above-mentioned information plus the uncertainties of $L_X$ ($\sigma_{Lx}$) are given. In this case, we use the bolometric X-ray luminosity $L_X$ emitted from within $r_{500}$.
We excluded two clusters that appear to have a "negative" upper-limit $L_X$ uncertainty ($L_{X,max}<L_X$). For these 364 galaxy clusters, the median relative uncertainties for $L_X$ and $T$ are $\left(\dfrac{\sigma_{Lx}}{L_X}\right)=40.7\%$
 and $\left(\dfrac{\sigma_T}{T}\right)=21.1\%,$ respectively, which are considerably larger than the relative $T-$uncertainty of ACC. Most of these clusters (214, $\sim 60\%$) have a spectroscopically determined redshift ($\sim 10$\% of them are X-ray redshifts) while the remaining 150 have photometrically determined redshifts. The redshift uncertainty was assumed to be zero during the derivation of $L_X$ \citep{lloyd}.

From these, only three are already contained in the ACC sample, making the two catalogs almost independent. The cosmological parameters used to derive $L_X$ were $H_0=70\  \text{km}\  \text{s}^{-1} \text{Mpc}^{-1},\ \Omega_{\text{m}}=0.3,$ and $\Omega_{\Lambda}=0.7$.

\section{Analysis method} \label{analysis_method}

As previously stated, we use a new method to try to identify anisotropies of the extragalactic sky properties using the $L_X-T$ relation. The fact that $L_X$ heavily depends on the cosmological parameters through the luminosity distance, and that $T$ can be measured regardless of the cosmology, is of crucial importance.  

\subsection{Form of the X-ray luminosity-temperature relation}

The power-law form of the $L_X-T$ scaling relation that we use, following Mittal et al. (2011), is given by

\begin{equation}
\frac{L_X}{10^{44}\ \text{erg/s}}\ E(z)^{-1}=a\times \left(\frac{T}{4\ \text{keV}}\right)^b.
\label{eq1}
\end{equation}

The parameter fitting is performed using the  $\chi^2$-minimization method, using the logarithmic form of the $L_X-T$ relation. Additionally, we use the reversed $T-L_X$ relation as well when the $\sigma_T$ values are larger than the $\sigma_{Lx}$ ones. Both of these cases are displayed in Eq. (\ref{eq2}).

\begin{equation}
\begin{aligned}
\log{L'_X}&=\log{a}+b\log{T'}  \\[0.5cm]
\log{T'}&=\frac{\log{L'_X}-\log{a}}{b}.
\label{eq2}
\end{aligned}
\end{equation}

\subsection{Fitting procedure}\label{fitting}

The constraining of $a$ and $b$ is performed by minimizing $\chi^2$ as displayed in Eq. (\ref{eq3}) for $L_X-T$ or $T-L_X$ fitting,  respectively, for $N$ data  \footnote{We set as $L'_X=\dfrac{L_X}{10^{44}\ \text{erg/s}}\ E(z)^{-1}$  and $T'=\dfrac{T}{4\ \text{keV}}$.}:

\begin{equation}
\chi^2_L=\sum\limits_{i=1}^N\left(\frac{\log{(L'_{X,obs})}-\log{[L'_{X,th}(T',\mathbf{p})]}}{\sigma _{\log{L},i}}\right)^2  
\label{eq3}
,\end{equation}

\begin{equation}
\chi^2_T=\sum\limits_{i=1}^N\left(\frac{\log{(T'_{obs})}-\log{[T'_{th}(L'_X,\mathbf{p})]}}{\sigma _{\log{T,i}}}\right)^2,\end{equation}

where the numerator represents the difference between the measured value of $L'_X$ (or $T'$) with respect to the theoretical prediction of the quantity based on Eq. (\ref{eq2}).  The expected value is based on the measured value of $T'$ (or $L'_X$) of the cluster as well as on the free parameters $\mathbf{p}$. Also, $\sigma _{\log{L,T,i}}$ are the  Gaussian logarithmic uncertainties which are derived as indicated in \citet{reiprich} \footnote{$\sigma_{\log{x}}=\frac{x^+-x^-}{2x}$, where $x^+$ and $x^-$ are the upper and lower limits of the main value $x$ of a quantity, considering its uncertainty.}. The derivation of the $3\sigma$ uncertainties that are displayed with every best-fit value, is based on the usual $\Delta\chi^2=\chi^2-\chi^2_{min}$ limits ($\Delta\chi^2\leq 9$ or 11.83 for 1 or 2 fitted parameters, respectively).

In order to account for the uncertainties of the data in both axes, we consider a purely geometrical reasoning. Firstly, we perform the fitting considering only the y-axis uncertainties, obtaining the best-fit value for the slope, $b_1$. Then, we project the x-axis uncertainty to the y-axis, adding it to the already existing uncertainty of the y-axis quantity. For the $L_X-T$ case, this reads as $\sigma^2_{\log{L_X},\text{final}}=\sigma^2_{\log{L_X},\text{initial}}+(b_1\times \sigma_{\log{T}})^2$, while for $T-L_X$ fitting, we have $\sigma^2_{\log{T},\text{final}}=\sigma^2_{\log{T},\text{initial}}+\left(\dfrac{\sigma_{\log{Lx}}}{b_1}\right)^2$. Eventually, we repeat the procedure with the new y-uncertainties to obtain the final constraints of $a$ and $b$ of Eq. (\ref{eq1}).  This fitting method is equivalent to the one described by \citet{akritas} and  has been used by several studies; for example \citet{zhang}. Moreover, a  $\sim 100\%$ change of $b_1$ would cause a $\sim 6\%$ shift of the final best-fit values, suggesting that the obtained results are not sensitive to the uncertainty conversion that we apply.  

\subsection{Identification of anisotropies} \label{identif_anis}

In order to pinpoint the solid angles in the sky that seem to share the largest deviation between them, we "scan" the sky as following: 
We consider the sky region with the Galactic coordinates $l\in [-45^{\circ},+45^{\circ}]$ (here we avoid the notation $l\in [315^{\circ},45^{\circ}]$) and $b\in [-90^{\circ},+90^{\circ}]$. We obtain the best-fit value for the 
fitted parameters.

Subsequently, we shift this region by $5^{\circ}$ towards larger values of $l$ (keeping the same size $\Delta l=90^{\circ}$), obtaining again the best-fit values. We repeat until the entire sky is scanned, returning to the initial position of the sky region. Each region contains $19\%-30\%$ of all the clusters for ACC and  $14\%-43\%$ for XCS-DR1. 

Moreover, we follow the same steps to scan the sky in terms of the Galactic latitude $b$, where we consider regions with a width of $\Delta b=40^{\circ}$, with a shift of $10^{\circ}$ every time. The fewer clusters there are in a region, the larger the uncertainty in the derived result.
We do not use smaller solid angles, since they contain fewer data and eventually they are heavily affected by individual clusters that can act as outliers, making the results untrustworthy. 

\subsection{Statistical significance and outliers} \label{statistics}

We need a valid expression for the statistical significance of the results that can be applied to both cluster samples. For this purpose, we use the Bootstrap resampling method. In detail, we consider the remaining sample, after excluding the clusters contained in the sky region, for which we want to express the statistical significance of its best-fit results. Then, we randomly draw 10000 different groups of clusters with the same number of data as the excluded one, fitting the parameters we are interested in for each group. Consequently, we obtain the mean values and the standard deviation of the results, allowing us to express the frequency with which the best-fit values of the excluded subsample could randomly appear.

If we identify inconsistencies between different subsamples, we need to clarify whether or not this inconsistency is caused by certain outliers; to do this we exclude them and check to see if the inconsistency disappears. To this end, we apply the Jackknife resampling method to identify such possible data with strong effects in the final best-fit solutions. The procedure we follow is similar to that described in \citet{migkas}. For a given subsample with $N$ galaxy clusters, we exclude one cluster each time, calculating the best-fit values of the parameters of our choice. Thus, we obtain $N$ different best-fit values for $N$ different subsamples, containing $N-1$ clusters each time. If the best-fit values do not change significantly regardless of the excluded cluster, then we conclude that the peculiar behavior of the subsample is systematic and not caused by individual data.

Finally, we highlight the pair of independent solid angles with the largest tension, that does not depend on only a few cluster measurements.

\section{Results} 

\subsection{ACC}
\begin{figure}[hbtp]
               \includegraphics[width=0.49\textwidth, height=6cm]{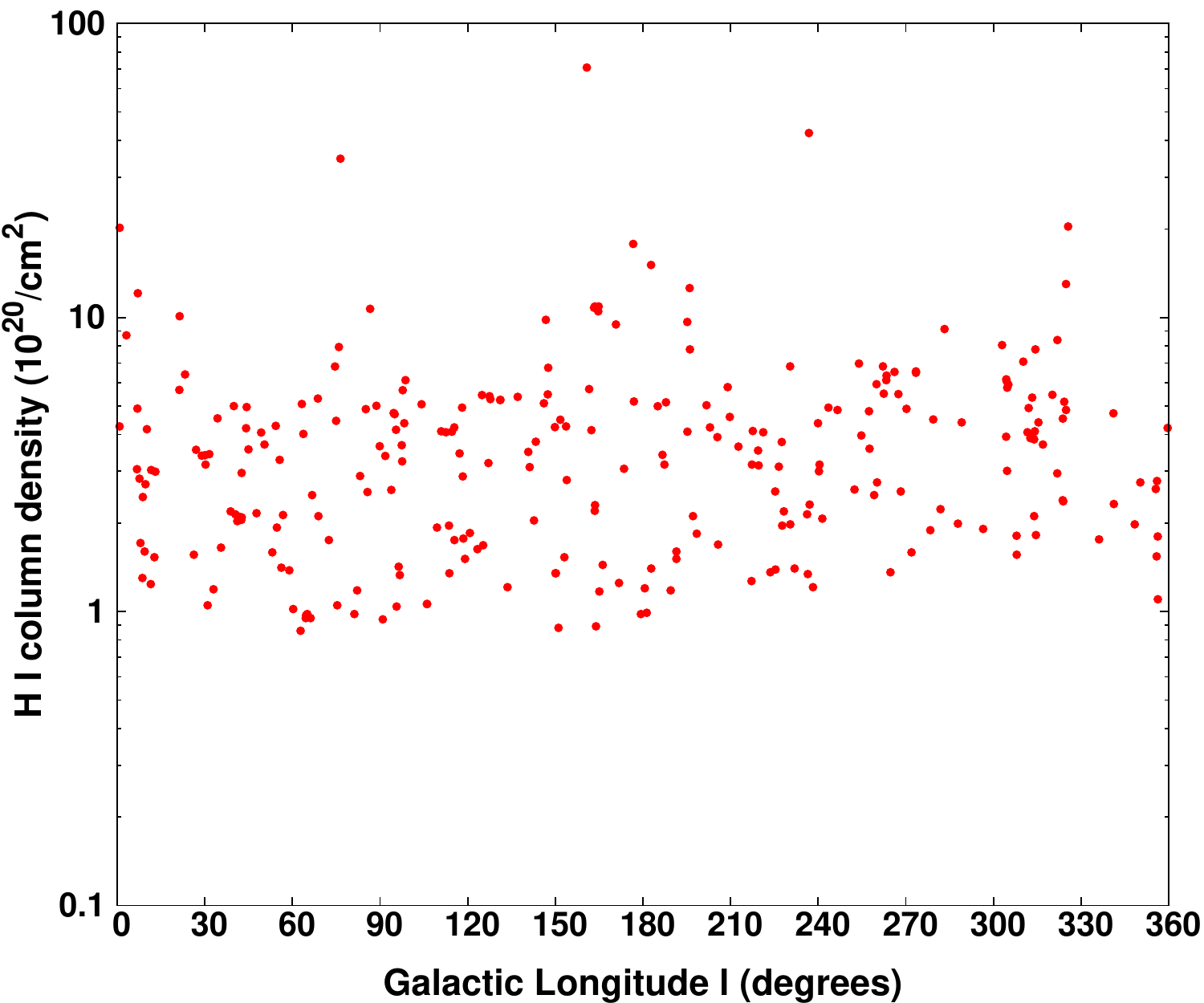}
               \includegraphics[width=0.49\textwidth, height=6cm]{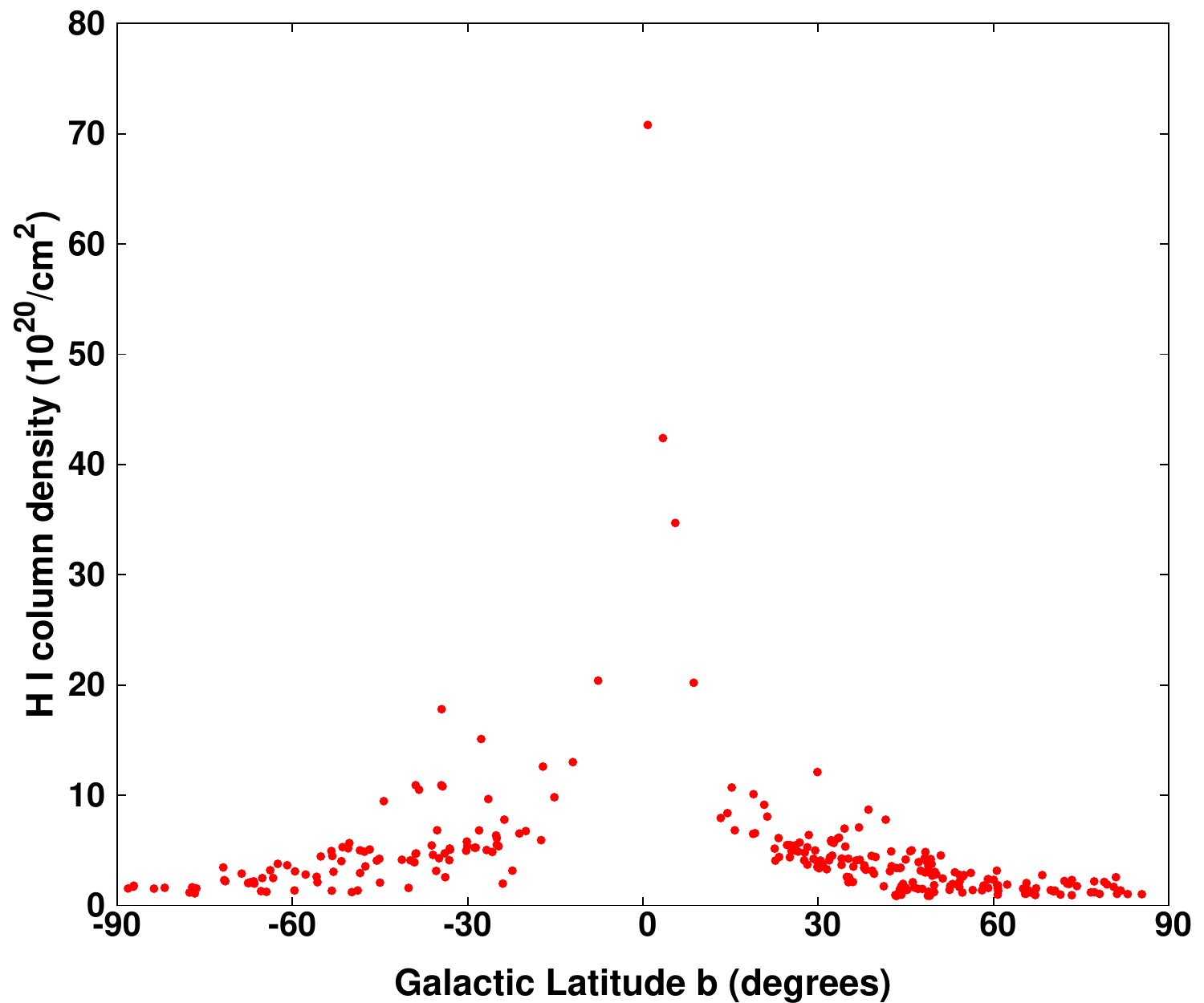}
               \caption{HI column density as given by \citet{dickey}, as a function of the Galactic longitude (top) and latitude (bottom).}
        \label{NH}
\end{figure}

As explained in Section \ref{acc_sample}, in order to correct the X-ray luminosities of 230 objects (out of 272 in total), \citet{horner} used the HI column density values ($N_{\text{HI}}$) as given by \citet{dickey}, based on 21cm measurements. The X-ray luminosities for the last 42 objects were corrected by us using the same method. As shown in Fig. \ref{NH}, $N_{HI}$ promptly increases for the Galactic latitudes $b\leq |20^{\circ}|$, while it appears to have a mild structure with respect to the Galactic longitude $l$ (small peaks every $\Delta l\sim 90^{\circ}$). This of course also depends on the $b-$distribution of the clusters for every $l$ region.

However, \citet{baumgartner} and \citet{schellenberger} have showed that the X-ray/total hydrogen column density ($N_{\text{H,tot}}$) roughly doubles compared to the HI column density, for $N_{\text{HI}}>10^{21}\ \text{cm}^{-2}$. If this increase of the absorption is not taken into account, it could lead to an underestimation of the X-ray luminosities of clusters near the Galactic plane as well as towards any direction with a HI column density of $N_{\text{HI}}>10^{21}\ \text{cm}^{-2}$. In order to obtain the correct $N_{\text{H,tot}}$ values for the sky position of every of these clusters, we use an online tool\footnote{\url{http://www.swift.ac.uk/analysis/nhtot/index.php}} which uses the method of \citet{willingale}. Finally, using \textsc{XSPEC}, we correct the $L_X$ values for all the respective clusters, prior to our analysis.

\subsubsection{General solution}

Since ACC does not come with $L_X$ uncertainties, we use the $T-L_X$ fitting procedure, based on Eq. (\ref{eq4}). However, the initially obtained normalization is not representative of the sample. This is due to the "overfitting" of some clusters with large $L_X$ and very low $T-$uncertainties that dominate the $\chi^2$-fit.

With the purpose of applying a more realistic approach, we insert an extra 5\% $L_X$ uncertainty to every data point, converting it to a $y-$axis uncertainty as described in Section \ref{analysis_method}, with $b_1=3.405$ (the best-fit value considering only $\sigma_T$). We choose this value for  $\sigma_{Lx}$ because we want it to be small compared to the median $\sigma_T$; additionally, it is the minimum value of inserted $\sigma_{Lx}$ that does not affect our result significantly, but allows us to obtain a more representative fitting of the data. Compared to the case with only $\sigma_{T}$, the slope changes by $\sim 1\%$, the normalization by $\sim 15\%$ and the reduced $\chi^2_{red}=\chi^2_{min}/\text{d.o.f.}$ by $\sim 50\%$. To ensure that our method does not bias our results, we always check whether or not the behaviors that we identify during the analysis exist also for the $\sigma_{Lx}=0$ case. We should note that \citet{horner} uses an inserted 20\% uncertainty for $L_X$ for all clusters, in order to constrain the $L_X-T$, stating that this has a minimum effect on the results. He also argues that the best-fit results do not significantly depend on the fitting method. We consider the case of the 20\% $L_X$ uncertainty as well, every time we have an interesting finding, in order to see how (and if) it is affected by the different value of $\sigma_{Lx}$.

The best-fit values with their $3\sigma$ credibility intervals for the 272 objects of ACC that we use are:

\begin{equation}
a=3.631^{+0.085}_{-0.083}\  \text{and}\ \   b=3.375^{+0.050}_{-0.045}\ \  \text{with}\ \  \chi^2_{min}/\text{d.o.f.}=38.606
.\end{equation} 

While the normalization value is considered to be typical, the slope is slightly large for typical galaxy cluster samples. Using the same sample, \citet{horner} found $b=3.49\pm 0.1$, fully consistent with our result, despite the different fitting procedures. \citet{fukazawa} also considered some massive elliptical galaxies along with the galaxy clusters and groups of ASCA, finding $b=3.17\pm 0.15$ for objects with a gas temperature of $1.5-15$ keV and  $b=3.74\pm 0.72$ for $1.5-5$ keV (90\% C.I.), which is again consistent with our derived value. Generally, the slope obtained by the $T-L_X$ fitting is expected to be somewhat different from the corresponding $L_X-T$ fitting value. Furthermore, the best-fit values of the $3\sigma$ uncertainties we recover are quite small due to the considerably large $\chi^2_{red}$. The latter emerges because the $L_X-T$ relation has significant intrinsic scatter and because the statistical uncertainty of $L_X$ is likely underestimated. \\[1cm]

\subsubsection{Different sky solid angles} \label{sky_angles}

The main goal of this project is to test the isotropy of the $L_X-T$ scaling relation. Before we apply the method described in Subsection \ref{identif_anis}, we divide the sky into hemispheres and derive the best-fit values for $a$ and $b$. The results are shown in Table \ref{tab2}.

\begin{table*}[hbtp]
\caption{\small{Best-fit values of the fitted parameters with their 3$\sigma$ credibility ranges for the four Galactic hemispheres.}}
\label{tab2}
\begin{center}
\begin{tabular}{ c  c  c  c}
\hline \hline

Hemisphere (No. of members)  & $a$ & $b$ & $\chi^2_{min}/\text{d.o.f.}$\\
\hline \hline

Northern (175) &\ $3.508^{+0.123}_{-0.080}$\ & \ $3.315^{+0.055}_{-0.060}$ \ & $38.614$\\ 
Southern (97)  &\ $3.890^{+0.137}_{-0.175}$\ & \ $3.520\pm 0.091$\ & 38.733\\ \hline
Eastern (150)  &\ $3.981^{+0.140}_{-0.135}$\ & \ $3.345\pm 0.069$\ & 33.419\\ 
Western (122) &\ $3.273^{+0.115}_{-0.111}$\ &\ $3.365^\pm 0.065$\ & 44.438 \\ \hline
All (272) &\ $3.631^{+0.085}_{-0.083}$\ &\ $3.375^{+0.050}_{-0.045}$\ & 38.606\\
 \hline 
\end{tabular}
\end{center}
\end{table*}

The northern and southern hemispheres do not appear to have significant deviations. For ACC, we do not display the probability contours of the $a-b$ solution space, since the reduced $\chi_{red}^2\gg 1$ and as a result, the real uncertainties should be larger than the ones derived by the usual $\Delta\chi^2$ limits. On the other hand, for the second pair of hemispheres, it is easily seen that there is a modest apparent inconsistency in the normalization value. In addition,
it is noteworthy that the slope $b$ is similar for all the Galactic hemispheres except for the southern, which seems to have a slightly increased $b$ value. Since the southern Galactic hemisphere contains relatively few clusters, it is more easily affected by outliers. In fact, if we exclude its most extreme outlier, galaxy cluster 2A 0335+096 (we analytically explain why this is  so further below), its slope shifts to $b=3.405^{+0.090}_{-0.085}$ , becoming more consistent with the all the other Galactic hemispheres. Moreover, its normalization also shifts to $a=3.428^{+0.162}_{-0.116}$, becoming totally consistent with the northern Galactic hemisphere.

Furthermore, we are more interested in expressing any occurring deviations in terms of the normalization $a$, since the latter is more closely related to the cosmological parameters than to the slope (Section \ref{intro}). We are not interested in the true values of $a$ for every region, but in comparing the consistency between different sky patches. Since the obtained $a$ value for a subsample clearly depends on the corresponding $b$ value, we prefer to use the same slope value for every subsample so that the comparison is not biased. Even if two subsamples have significantly different slope values (being inconsistent in that manner) but similar $a$ values, this discrepancy will propagate in the $a$ values when we use the same slope for both, not allowing the deviation to be ignored. Thus, we fix the slope to $b=3.375$ and we only fit the normalization $a$ when we look for anisotropies with the "scanning" method.

\begin{figure*}[hbtp]
               \includegraphics[width=0.49\textwidth, height=6cm]{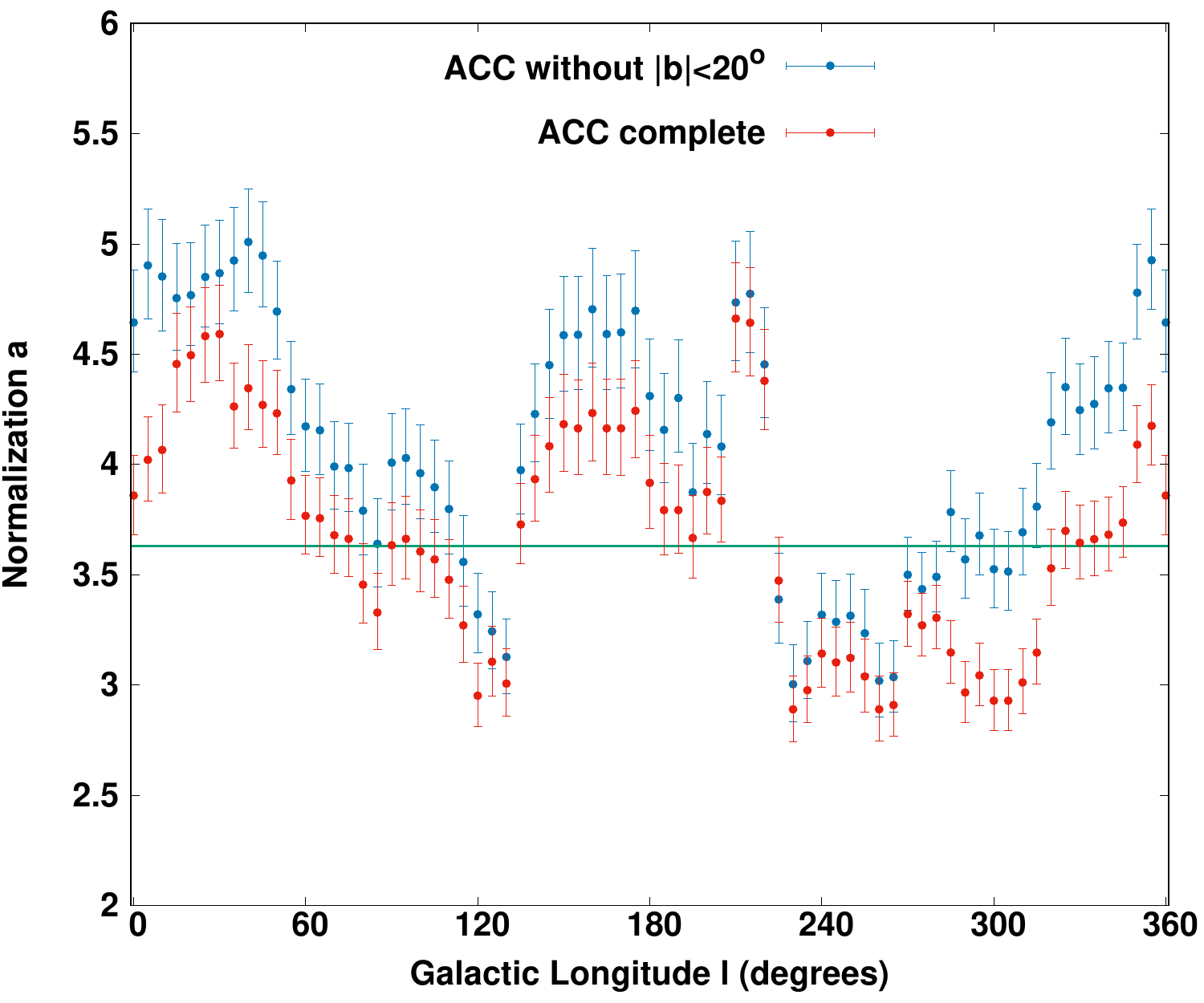}
               \includegraphics[width=0.49\textwidth, height=6cm]{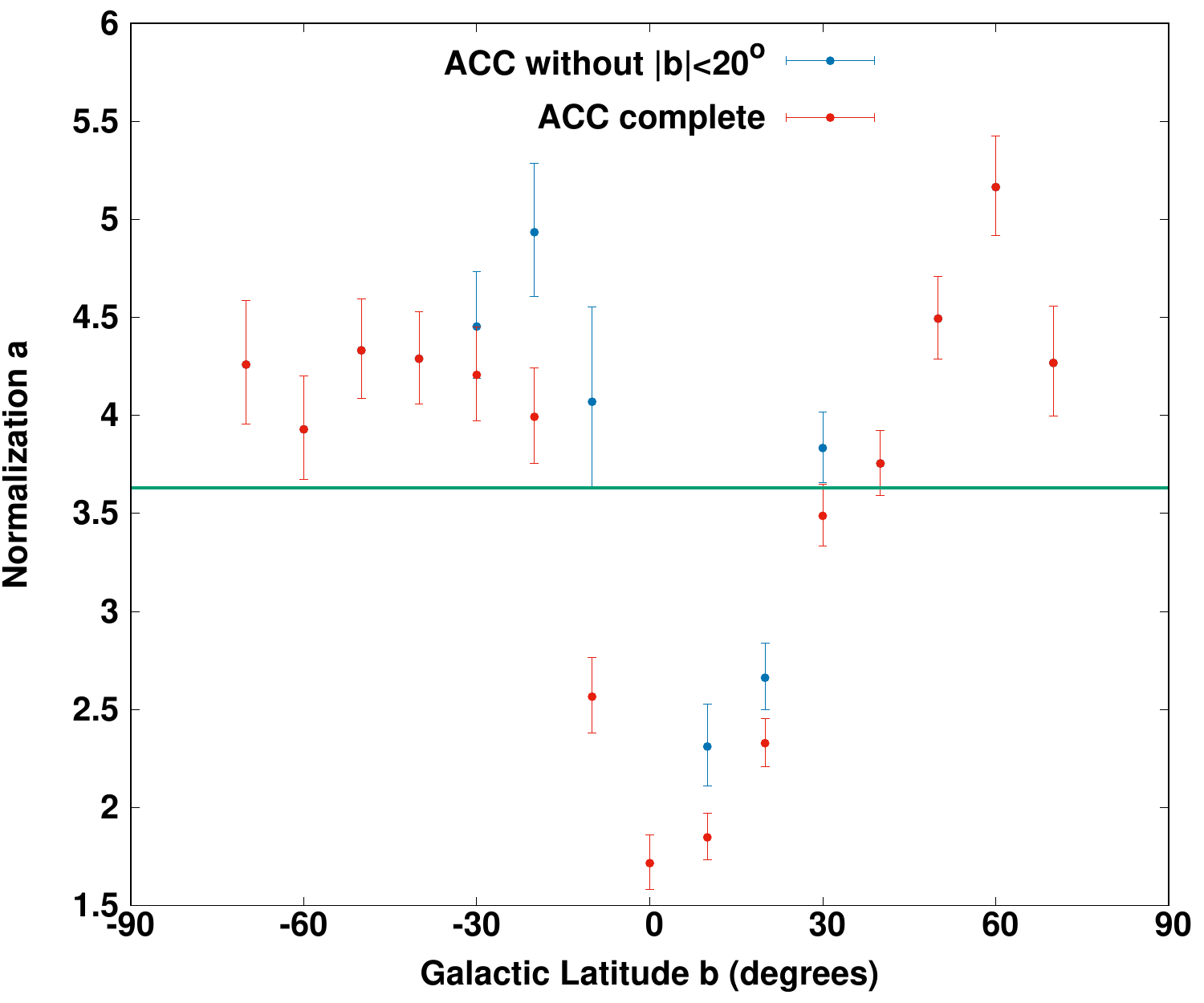}
               \caption{Best-fit value of the normalization for every sky region of ACC with: \textit{Left:} $\Delta l=90^{\circ}, \ \Delta b=180^{\circ}$ as a function of its central Galactic longitude and \textit{Right:} $\Delta l=360^{\circ}, \ \Delta b=40^{\circ}$ as a function of its central Galactic latitude. Given the bin widths, the data points are obviously strongly correlated.}
        \label{a-fit}
\end{figure*}

As displayed in the left panel of Fig. \ref{a-fit}, the normalization of the $L_X-T$ relation strongly fluctuates with the Galactic longitude, while surprisingly, the highest and lowest peaks are separated by $\Delta l\approx 90^{\circ}$. Another interesting feature is the smooth transition to the peaks for $l\sim 30^{\circ}$ and $l\sim 270^{\circ}$. This indicates that clusters in these regions have a systematic behavior towards higher/lower normalization values and they do not cause random fluctuations of the derived $a$ value. 

Additionally, from the right panel of Fig. \ref{a-fit}, we can see that the normalization heavily drops as we move towards the Galactic plane, suggesting that the X-ray luminosities of these low$-b$ clusters have likely not been properly derived. Here it is essential to remind the reader that we correct the $L_X$ values for the total hydrogen column density absorption, since the given $L_X$ are supposed to be corrected for the absorption only due to the neutral atomic hydrogen column density. This "dip" in the $a$ value is not relieved even for the most extreme case for which we assume that no absorption correction has been applied to the data (knowing this is not the case, since the catalog contains both the absorbed and the unabsorbed flux and luminosity values). Applying the correction ourselves (practically "overcorrecting" for the absorption twice in a row), the lowest point at $b=0^{\circ}\pm 20^{\circ}$ shifts to $a=2.187^{+0.125}_{-0.121}$). After using the Jackknife
method, we realize that this behavior cannot be  entirely attributed to outliers. Consequently, when looking for possible anisotropies, we have to be careful with the consequences of clusters with $|b|\leq 20^{\circ}$ in our subsamples. 
Additionally, we see that when we exclude all (16) clusters within $|b|\leq 20^{\circ}$, $a$ increases significantly for $b\in [-40^{\circ},30^{\circ}]$. 
The lowest $a$ in this case appears for $b\in [20^{\circ},30^{\circ}]$ (25 clusters), with $a=2.31\pm 0.209$. The error bars displayed  are obtained from the $\Delta\chi^{2}$ limits and since $\chi^2_{red}\gg 1$, they are not representative of the true uncertainties. 

In order to put a proper probabilistic value on the deviations, we use the Bootstrap method as described in Subsection \ref{statistics}, where we consider 10000 random subsamples every time, after we have excluded the subsample of interest. The $\sigma$ value in the parentheses henceforth represents the Gaussian deviation of the result as derived from the Bootstrap, with respect to the rest of the sample. According to this method, the region within $b\in [20^{\circ},30^{\circ}]$ is consistent at $\sim 2\sigma$ with the rest of the sample (excluding low-$b$ clusters) and therefore is not considered as statistically significant. However, when we use the entire sample, the 16 clusters within $|b|\leq 20^{\circ}$ have a $\sim 2.4\sigma$ deviation from the rest.

For the left panel of Fig. \ref{a-fit} again, the peaks at $l\sim 120^{\circ}$ and $l\sim 215^{\circ}$ have the maximum deviation between them, but due to the sudden transition of their values, it is indicated that this is caused by single clusters acting as outliers. In order to confirm this, we again use the Jackknife method. In fact, the normalization of the region with central $l\in [165^{\circ},260^{\circ}]$ (66 clusters) shifts from $a=4.596^{+0.248}_{-0.235}$ ($2.68\sigma$) to $a=3.812^{+0.217}_{-0.206}$ ($0.87\sigma$) when we exclude the galaxy cluster 2A 0335+096, making the behavior of the region  consistent with the rest of the sample. In particular, 2A 0335+096 has a temperature and a bolometric X-ray luminosity of $T=2.86\pm 0.03$ and $L_X=1.095\times 10^{45}$ erg/s, respectively. However, this specific cluster has been found to have a higher temperature by \citet{ikebe} ($T=3.64^{+0.09}_{-0.08}$ keV) and \citet{hudson} ($T=3.53^{+0.10}_{-0.13}$ keV), who both used a double thermal modeling with ASCA and Chandra \citep{chandra}, respectively, while \citet{horner} uses a single thermal modeling.

\begin{figure*}[hbtp]
              \includegraphics[width=0.49\textwidth, height=6cm]{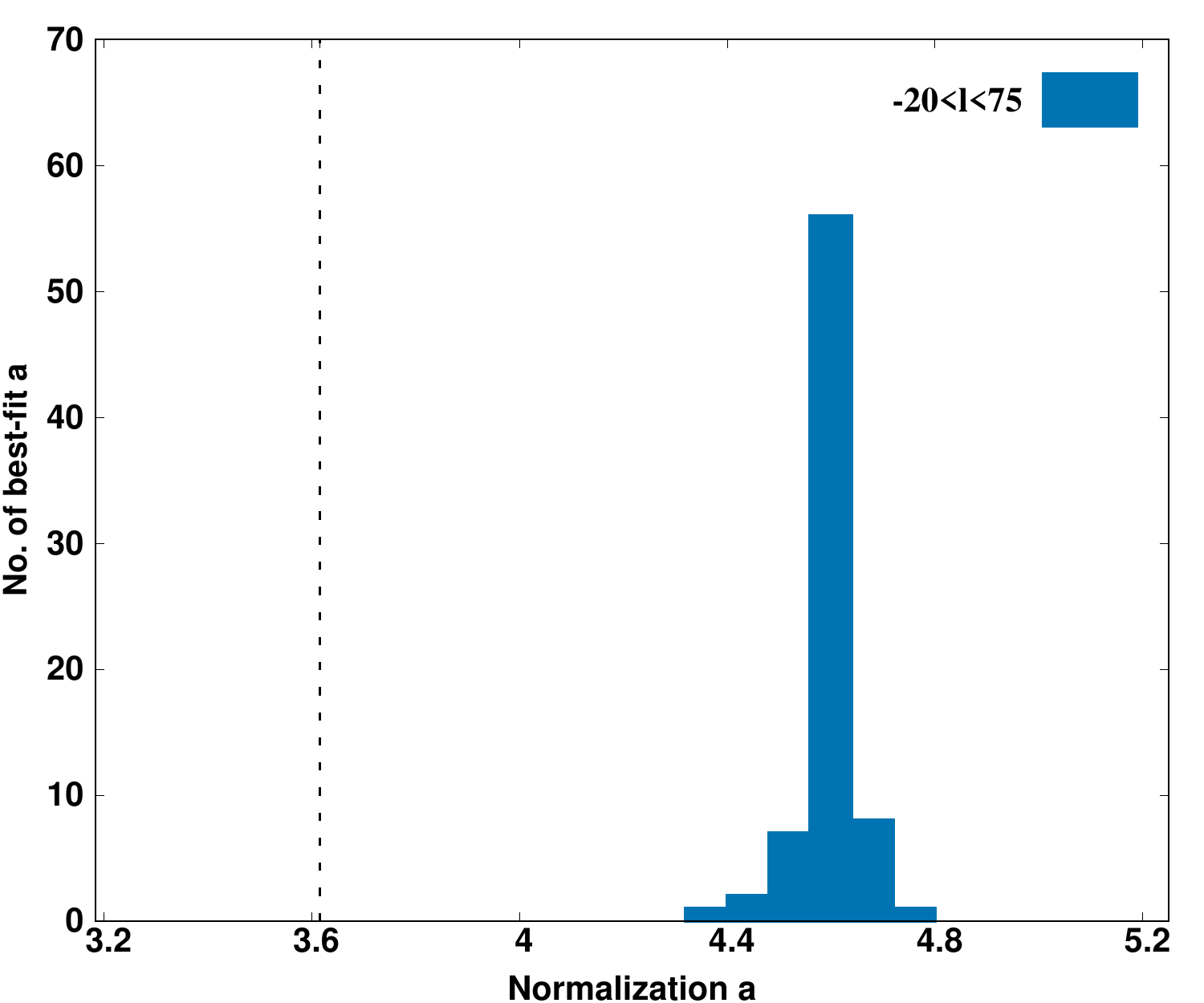}
              \includegraphics[width=0.49\textwidth, height=6cm]{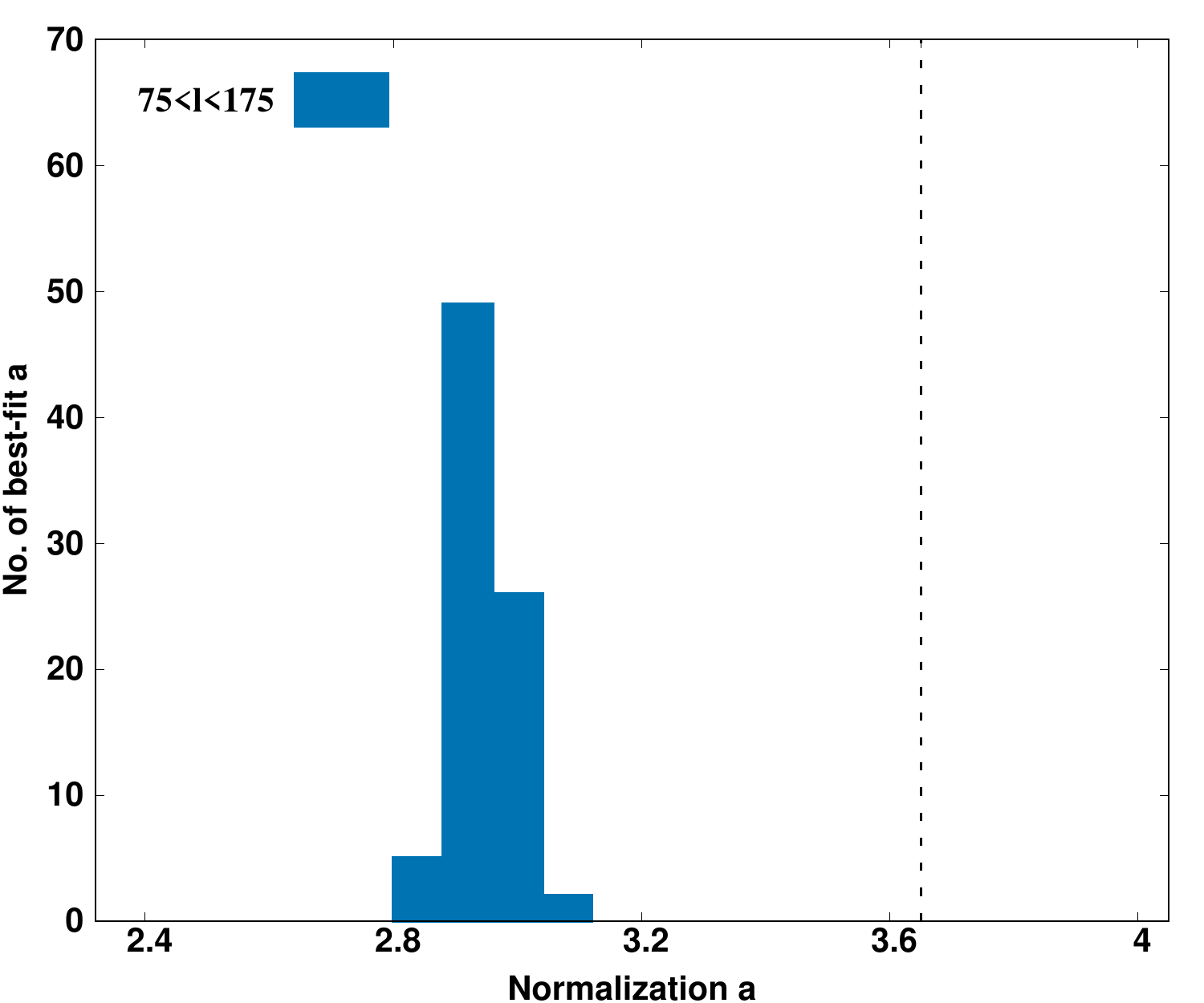}
              \includegraphics[width=0.49\textwidth, height=6cm]{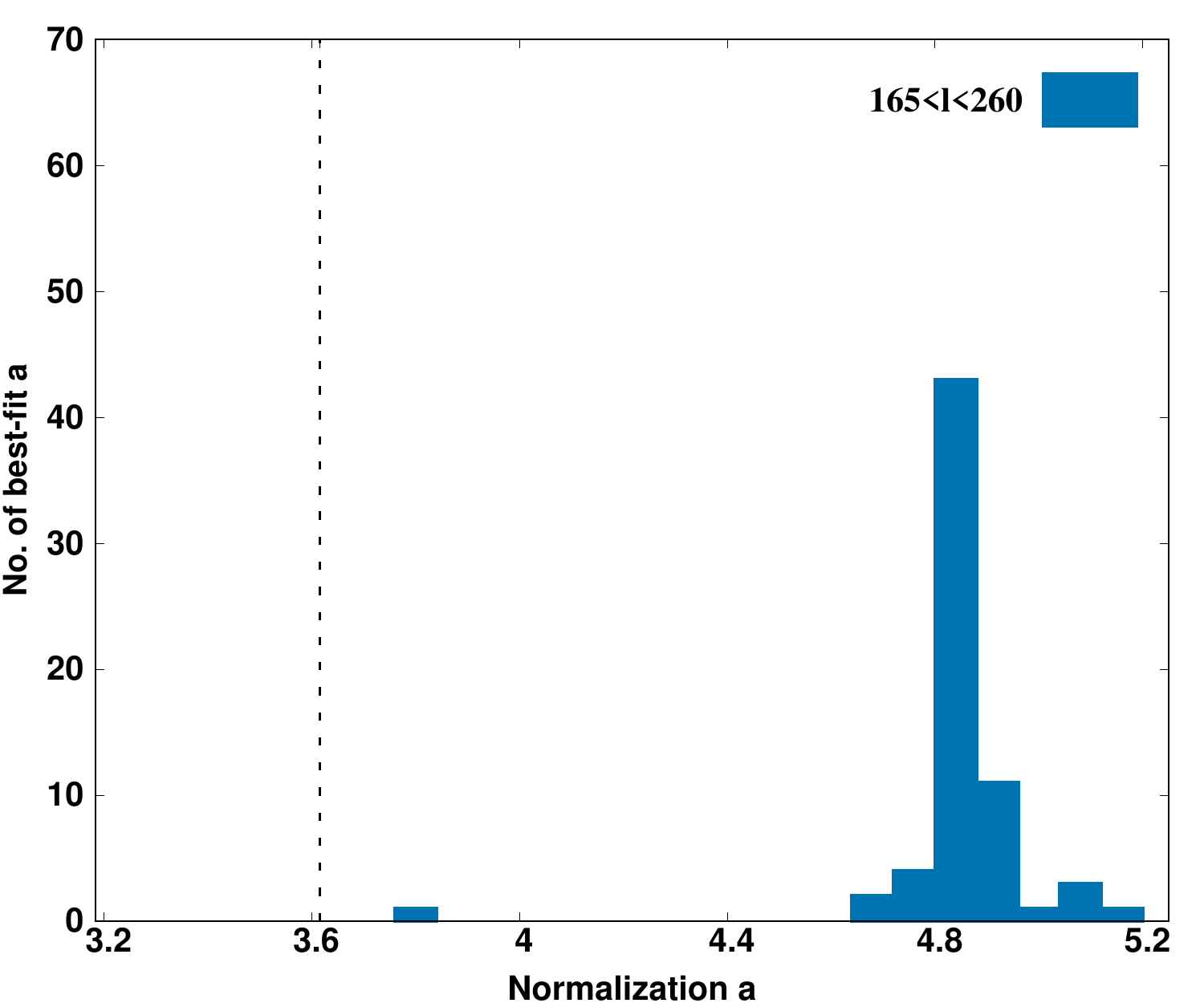}
             \includegraphics[width=0.49\textwidth, height=6cm]{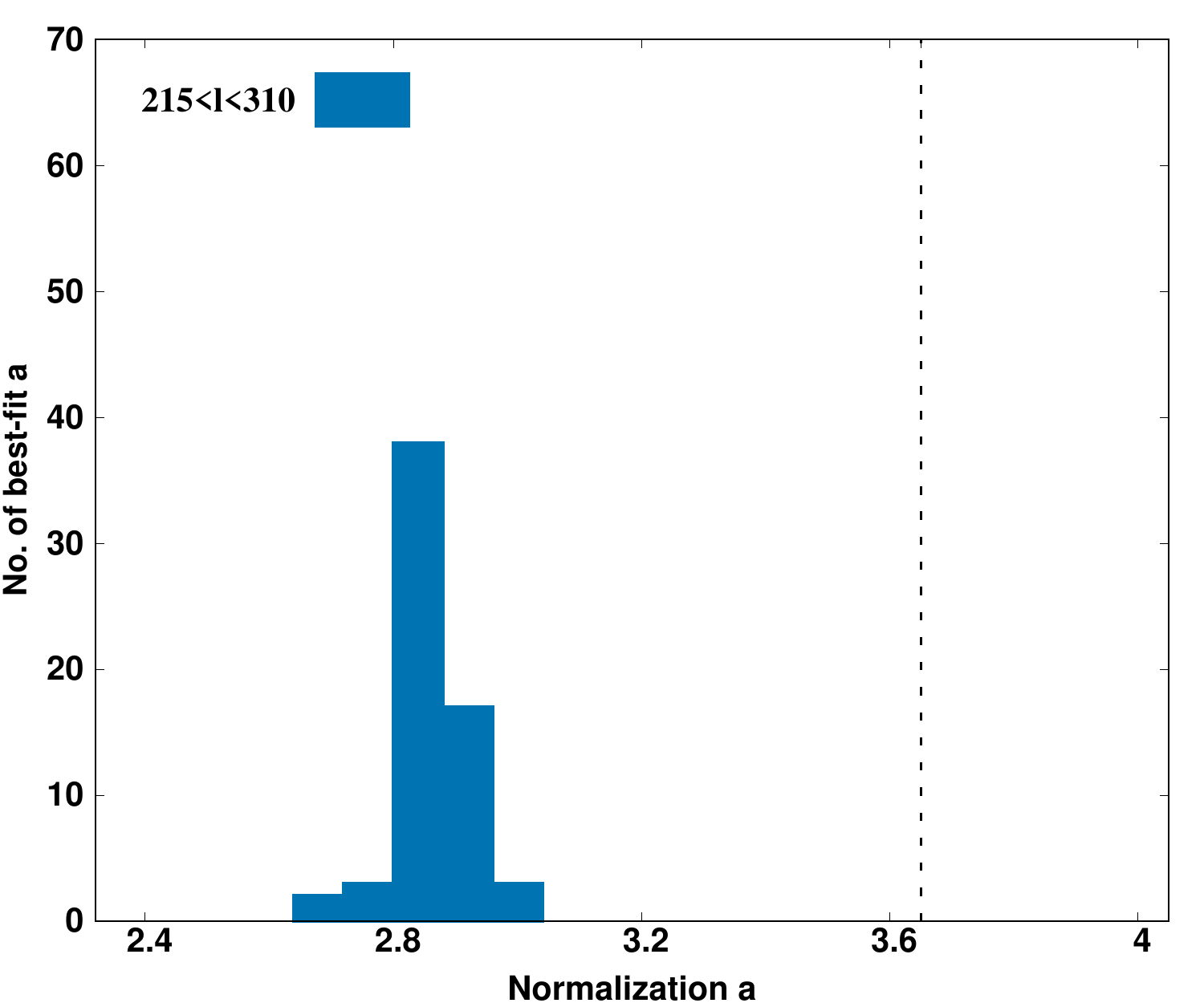}
         \caption{Distribution of the best-fit value of the normalization as obtained by the Jackknife resampling method, for the sky regions with 75 clusters within $l\in (-20^{\circ},75^{\circ})$ (top left), 82 clusters within $l\in (75^{\circ},175^{\circ})$ (top right), 66 clusters within $l\in (165^{\circ},260^{\circ})$ (bottom left) and 63 clusters within $l\in (215^{\circ},310^{\circ})$ (bottom right).}
        \label{jack}
\end{figure*}

Therefore, the region with practically the largest normalization is the one with $l\in [-20^{\circ},75^{\circ}]$ (75 clusters, where we consider as one the regions expressed by the two data points with the largest $a$), which has $a=4.563^{+0.216}_{-0.206}$ ($2.65 \sigma$). Applying the Jackknife, the minimum value $a=4.393^{+0.666}_{-0.619}$ ($2.28\sigma$) occurs when we exclude A2052 while the largest is $a=4.758^{+0.700}_{-0.654}$ ($3.10\sigma$), obtained after excluding  the Ophiuchus cluster. As shown in the top left panel of Fig. \ref{jack}, this region does not contain any significant outliers. Furthermore, if low$-b$ clusters are excluded from the whole sample, the deviation between $l\in [-20^{\circ},75^{\circ}]$ and the rest slightly decreases to $2.09\sigma$. From now on, we will call this region Group A. However, in this case the region with the maximum deviation from the rest of the sample is $l\in [-10^{\circ},90^{\circ}]$, with $a=4.932^{+0.232}_{-0.223}$ ($2.47\sigma$).

On the other hand, the lowest normalization is the one for the region with $l\in [75^{\circ},175^{\circ}]$ (82 clusters) which rises from $a=2.963^{+0.145}_{-0.137}$ ($2.06\sigma$) to $a=3.102^{+0.516}_{-0.468}$ ($1.77\sigma$) when we exclude the galaxy cluster 3C 129, which is not such a big effect. Furthermore, excluding only A1885, we obtain $a=2.830^{+0.476}_{-0.428}$ ($2.34\sigma$). If we entirely exclude clusters with $|b|\leq 20^{\circ}$, the normalization of the region changes to $a=3.312^{+0.179}_{-0.168}$ while the deviation of this region remains the same ($2.04\sigma$). We refer to this region as Group C. 

For the other low$-a$ region with $l\in [215^{\circ},310^{\circ}]$ which contains 63 clusters, $a$ shifts from $2.891^{+0.143}_{-0.136}$ ($1.86\sigma$) to $3.020^{+0.505}_{-0.455}$ ($1.65\sigma$) after excluding AS636. However, it also shifts to $2.713^{+0.458}_{-0.409}$ ($2.16\sigma$) when we only exclude the galaxy cluster PKS 0745-19. Excluding, once more, all the clusters with $|b|\leq 20^{\circ}$, we obtain $=3.007^{+0.162}_{-0.154}$ and the deviation of this region shifts to $2.36\sigma$, becoming less consistent with the general solution. From now on we refer to this region as Group B.

All these results show that there is a strong tension mainly between Group A and the rest of the sample, as well as between Group A and Groups B and C. If we exclude both Groups A and C from the sample and we apply the Bootstrap method, we find that the deviation of Group A from the rest of the sample is still at $2.21\sigma$, while Group C is at $2.07\sigma$. However, it is necessary to point out here that now the remaining sample, from which we draw the 10000 random subsamples of the same size as the Groups, consists of 115 clusters. Hence, the 75 or 82 clusters of Groups A and C, respectively, constitute a large fraction of the remaining 115. Therefore, the 10000 random subsamples will be highly correlated and this could lead to a decreased standard deviation of the Gaussian results, eventually leading to an overestimation of the tension between the Groups and the rest. Therefore, we must be cautious about the deviations that occur in the case where the size of the subsamples is $>50\%$ of the sample from where we draw them.

Moreover, since we are dealing with small subsamples with large scatter, we must be very careful with the effect that outliers have on the results. Consequently, we apply the Jackknife method again to the rest of the sample, finding that the biggest effect comes from the galaxy cluster 2A 0335+096, as it was for the sky region within $l\in (165^{\circ},260^{\circ})$. If we also exclude this cluster, the deviation for Group C heavily drops to $1.07\sigma$ while for Group A this increases to $3.71\sigma$. As expected, if we now compare Group A with the rest of the sample including Group C, the deviation is at $3.64\sigma$, much higher than the case where we include 2A 0335+096 in the rest of the sample. Accordingly, comparing Group C with the rest of the sample (including Group A) we find a tension of $1.90\sigma$, slightly lower than the original case.

Assuming a typical subsample size of 77 galaxy clusters and considering the whole sample, we obtain that the deviation between Groups A and C is $3.84\sigma$  \footnote{Here we divide the difference between the best-fit $a$ values for the two Groups with the standard deviation obtained by the random subsamples, $\dfrac{a_{\text{best},2}-a_{\text{best},1}}{\sigma}$}, which is statistically very significant, although we stress again our reservations for this result, based on the sizes of the samples used. If we take the most conservative values excluding the two most important clusters for each group (as we identified them from the Jackknife), the tension is still at $3.11\sigma$. Excluding 2A 0335+096, the deviation between Groups A and C becomes $4.75\sigma$, and $3.88\sigma$ when we again exclude the most extreme cluster of each Group. This increased tension is obtained due to the lower scatter of the best-fit values of the normalization when 2A 0335+096 is not included, leading to a lower standard deviation of the Gaussian results.

Excluding only Groups A and B this time and applying the Bootstrap to the rest of the sample, we find that Group B deviates by only $1.36\sigma$, while Group A deviates by $1.83\sigma$, mainly because the standard deviation is large. If we also exclude just 2A 0335+096, the deviation of Group A rapidly increases to $3.28\sigma$. All these firmly indicate the large effect of this particular galaxy cluster and how sensitive our results are to outliers when we consider small samples.

Additionally, we test the case with no inserted $5\%\  L_X$ uncertainty, but only accounting for the temperature uncertainties. The values of the normalization slightly change (no more than $15\%$). When we consider Group A and the rest of the sample, we obtain a deviation of $1.02\sigma$ which is relatively low. If we now exclude 2A 0335+096 from the rest of the sample the deviation boosts to $2.88\sigma$. The deviation values derived for Group C and the rest of the sample for these two cases are $1.72\sigma$ and $1.69\sigma$.

For the case where we exclude both Groups A and C from the sample and compare with the rest, the deviation for Group C slightly decreases to $1.81\sigma$. On the contrary, the deviation of Group A  drops from 2.21$\sigma$ to $0.78\sigma$, due to the very large standard deviation of the results (114\% larger). However, we have to consider the reason for which we initially inserted this extra uncertainty and this is to avoid the overfitting of some clusters with very low $\sigma_T$ that dominate the results. Thus, by only excluding the galaxy cluster 2A 0335+096 once more, the normalization of the rest of the sample drops by $28\%$ and the deviation of Group A rockets to $2.72\sigma$ while Group C is now at $1.27\sigma$. Consequently, we conclude that this small inserted luminosity uncertainty is necessary to derive trustworthy results, even if the deviation remains roughly the same without it.

Finally, we use an inserted $L_X$ uncertainty of 20\%, following \citet{horner}. Now, in the cases that we include or not 2A 0335+096 in the rest of the sample, Group A deviates by $3.42\sigma$ and $3.84\sigma,$ respectively. On the other hand, Group C deviates by $2.26\sigma$ and $2.14\sigma$ for these two cases, respectively. All these show that the apparent deviations between these Groups do not strongly depend on the inserted $L_X$ uncertainty.

\begin{figure}[hbtp]
                \includegraphics[width=0.49\textwidth, height=6cm]{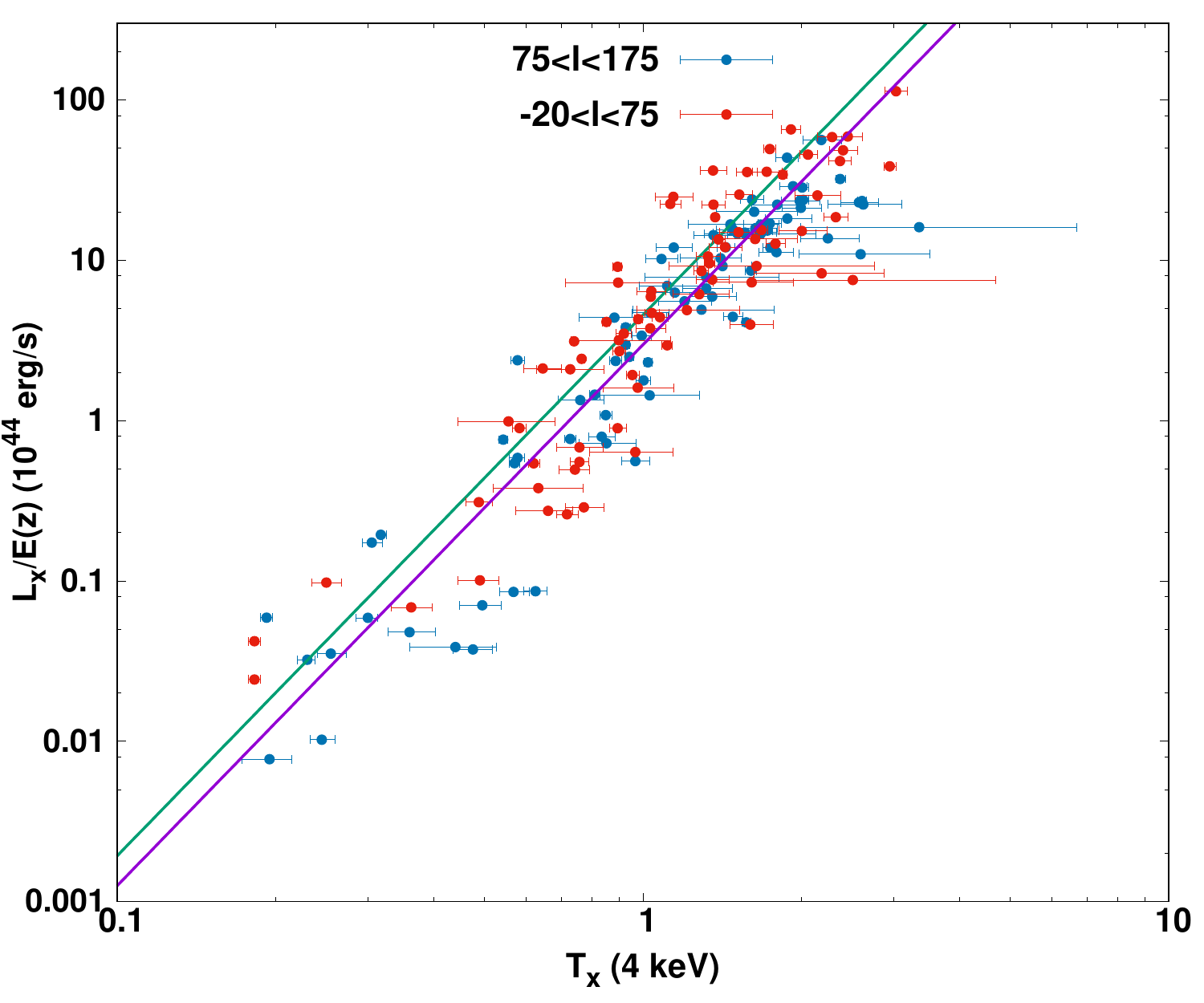}
        \caption{Bolometric luminosity $L_X$ as a function of temperature $T$ for the sky regions within $l\in (-20^{\circ},75^{\circ})$ (Group A, red) and $l\in (75^{\circ},175^{\circ})$ (Group C, blue). The best-fit functions (with a fixed slope of $b=3.375$) are also displayed with green for Group A ($a=4.563$) and with purple for Group C ($a=2.963$).}
        \label{L-T}
\end{figure}

In Fig. \ref{L-T}, the $L_X-T$ plane for Groups A and C is displayed.

\subsubsection{Possible causes} \label{causes}

\paragraph{Distributions of $T,\ z$ and $\sigma_{\log{Lx}}$.}

Since ACC does not have a specific selection function, it is necessary to test if these apparent anisotropies are caused by different temperature, redshift, or uncertainty distributions in the different sky regions. 

If the $L_X-T$ relation were not described well by a power-law for the whole temperature range (from groups to clusters), then having relatively more clusters with higher temperatures in one subsample compared to another could potentially bias the results. Similarly, if the redshift evolution of the $L_X-T$ relation were not satisfyingly described by the self-similar $E(z)$ factor, more high-$z$ data in one subsample than in another one would also add a bias in our derived normalization values.

\begin{figure*}[hbtp]
       \includegraphics[width=0.32\textwidth, height=5cm]{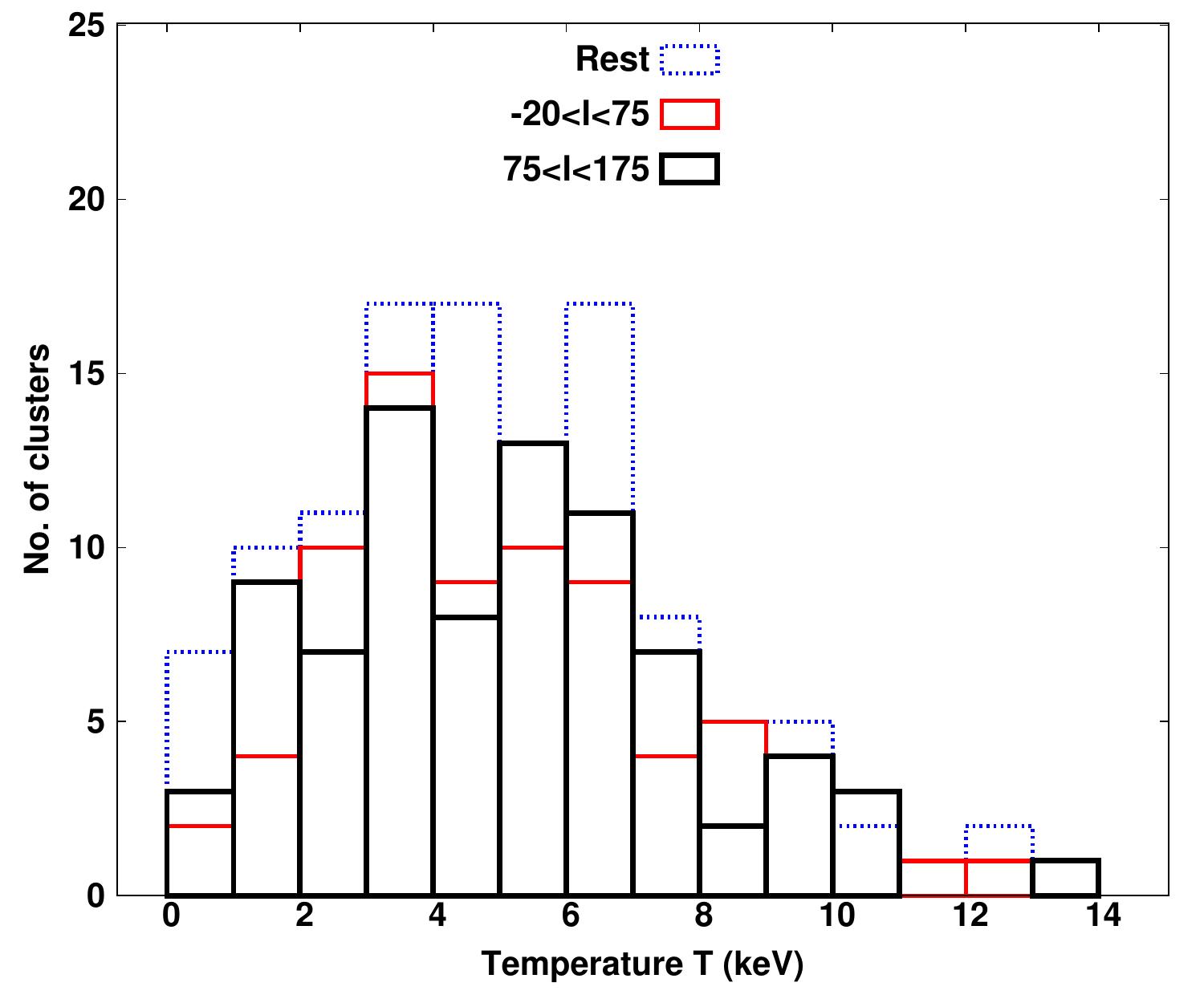}
       \includegraphics[width=0.32\textwidth, height=5cm]{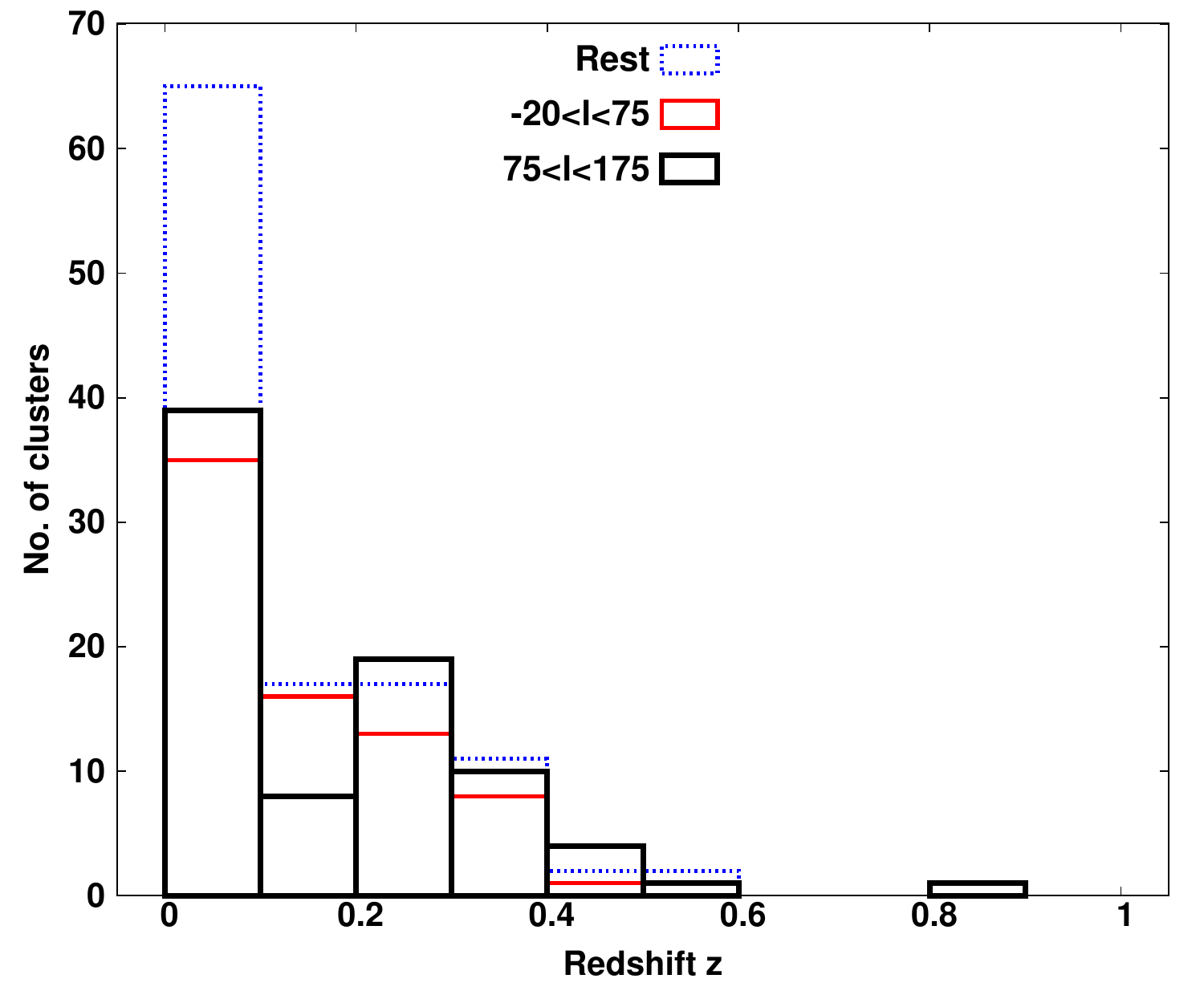}
      \includegraphics[width=0.32\textwidth, height=5cm]{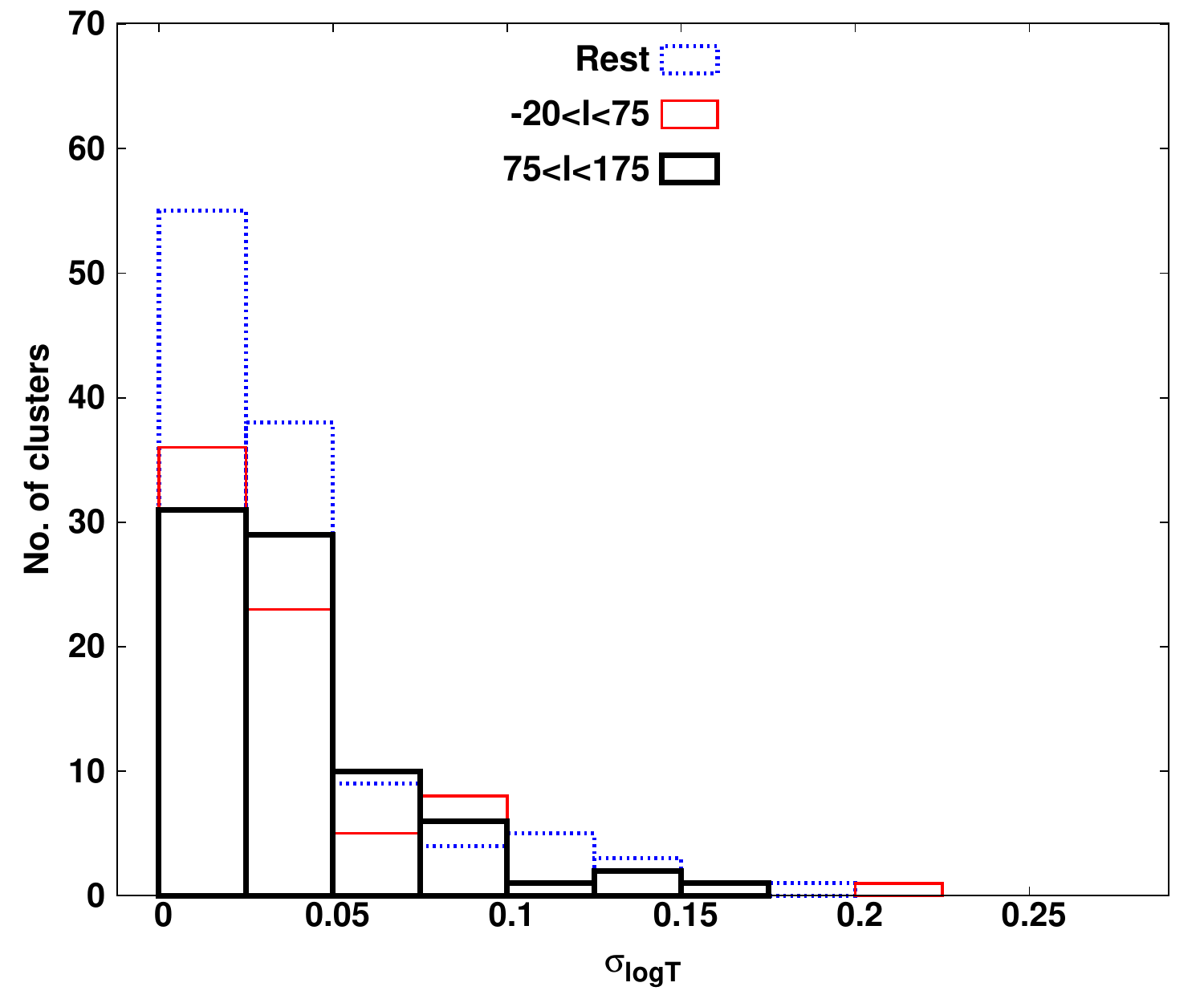}
       \caption{Distribution of the temperature (left), redshift (center) and temperature uncertainties (right) of the galaxy clusters contained in Group A (red), Group C (black) and the rest of the sample (blue).}
        \label{distr}
\end{figure*}

In Fig. \ref{distr}, these distributions for Groups A, C, and the rest of the sample are displayed.

The temperature and the uncertainty distributions of the three subsamples are very similar. Furthermore, the redshift distributions of Groups A and C are comparable, while the rest of the sample has a higher fraction of low$-z$ clusters than the two groups. However, this does not seem to play any role, since the two groups with similar distributions have such different best-fit values. This implies that the observed deviation is not the result of different selection effects.

In order to further investigate the reason for the behavior of the two different sky regions, we compare the best-fit normalization values of these three subsamples as they occur for low and high $T$ and $z$. The results are shown in Table \ref{subsamples-asca}.

\begin{table*}[hbtp]
\caption{\small{Best-fit normalization values with their 3$\sigma$ credibility ranges for Group A, Group C, and the rest of the sample of the ACC sample, for low and high temperature and redshift ranges.}}
\label{subsamples-asca}
\begin{center}
\begin{tabular}{ c  c  c  c  c  c}
\hline \hline

Subsample (No. of members)  & All & $T\leq 5$ keV & $T>5$ keV & $z\leq 0.1$ & $z>0.1$\\
\hline \hline

Group A (75) &\ $4.563^{+0.216}_{-0.206}$\ & \ $4.970^{+0.304}_{-0.287}$ \ & $3.924^{+0.311}_{-0.288}$ & $4.574^{+0.245}_{-0.232}$ & $4.422^{+0.502}_{-0.450}$\\ 
Group C (82) & $2.963^{+0.145}_{-0.137}$ & \ $3.687^{+0.237}_{-0.223}$\ & \ $2.141^{+0.169}_{-0.156}$ \ & $2.927^{+0.156}_{-0.148}$ & $3.182^{+0.444}_{-0.391}$ \\ \hline
Rest (115) &\ $3.578^{+0.132}_{-0.127}$\ & \ $4.261^{+0.197}_{-0.189}$\ & $2.599^{+0.164}_{-0.154}$ & $3.411^{+0.137}_{-0.132}$ & $4.683^{+0.460}_{-0.421}$\\
 \hline 
\end{tabular}
\end{center}
\end{table*}

The tension between the Groups A, C, and the rest of the sample persists in both the low- and high$-T$ regimes. The relatively high value of the rest of the sample for the low$-T$ data is mainly caused by 2A 0335+096 and without it, it decreases to $a=3.729\pm 0.175$. Therefore, the strong deviation seems to be consistent for all temperatures. 

In the high-$z$ regime, there is only a small discrepancy in $a$ between the two Groups, while the rest of the sample has a larger $a$ than Group A. In the low-$z$ data however, the discrepancy is $\sim 3\sigma$. We have to take into account though that the large uncertainties of the derived results do not allow us to draw robust conclusions for the galaxy clusters with $z>0.1$. Moreover, since the number of data is very small for these $T$ and $z$ ranges, the best-fit values are very sensitive to the exact limits of the different ranges. Nevertheless, one could conclude that the apparent deviation seems to be stronger in the local Universe, something that could potentially be attributed to differences in the local structure, such as the existence of superclusters. This will be further investigated for the XCS-DR1 sample, as described in Section \ref{causes-XCS}.
In addition, clusters with $|b|\leq 20^{\circ}$ do not change the results of the rest of the sample to an important degree, while excluding them shifts $a$ for Group C ($z\leq 0.1$) to $a=2.324^{+0.426}_{-0.376}$ and for Group A to $a=4.912^{+0.268}_{-0.277}$.

\paragraph{Redshift correction to the CMB frame and peculiar velocities.}

Another test is to check if the (conservatively assumed) heliocentric redshifts of the data affect the final normalization and if we need to use the redshifts of the clusters with respect to the CMB frame. 
To this end, we properly convert the redshifts of all the clusters in a way so to account for a bulk velocity of $371$ km/s towards $(l,b)\sim (270^{\circ}, 35^{\circ})$. The direction of our bulk motion within the CMB frame has been previously found by \citet{fixsen} to be towards $(l,b)\sim (264.14^{\circ}\pm 0.30^{\circ}, 48.26^{\circ}\pm 0.30^{\circ})$ with a velocity of $371\pm 1$ km/s. Also, \citet{bennett03} found the same values for the direction and velocity and \citet{watkins} found a bulk velocity of $407\pm 81$ km/s towards $(l,b)\sim (287^{\circ}\pm 9^{\circ}, 8^{\circ}\pm 6^{\circ})$. Thus, we adopt values for the direction in between the results of these three studies. We should note that the obtained results are not at all sensitive to the exact direction and amplitude of the relative motion.

Such a motion could lead us to observe smaller redshifts in this direction than the ones only due to the Hubble flow. Consequently, an underestimation of the distance and equivalently of the luminosity would take place, resulting in a lower normalization value.

Repeating the analysis up to now considering the CMB redshift correction, we see that this does not significantly affect the apparent deviation. The incompatibility between Groups A and C becomes somewhat smaller ($4.16\sigma$), while the one between Group A and the rest of the sample also slightly decreases ($2.56\sigma$). On the other hand, Group C has now a deviation of $2.53\sigma$ with the rest of the sample. Finally, it is remarkable that by only excluding galaxy cluster 2A 0335+096, the deviation of Group A becomes $3.45\sigma$ while the one of Group C remains almost constant.

In order for the tension between Group A and the rest of the sample to drop below 2$\sigma$, the velocity of our bulk motion towards $(l,b)\sim (270^{\circ}, 35^{\circ})$ would need to be $u_{bulk}\gtrsim 3030\ \text{km/s}$, which is obviously out of the question. The lowest possible bulk velocity that could decrease the above mentioned tension by same level  is $u_{bulk}\gtrsim 415\ \text{km/s}$ towards  $(l,b)\sim (210^{\circ}, 0^{\circ})$ (the antipodal point of the center of Group A in the sky). However, all these cases would induce a dipole-like apparent anisotropy with the normalization maxima and minima separated by $\sim180^{\circ}$,  and this does not seem to be the case, as shown in the left panel of Fig. \ref{a-fit}.

The obtained redshifts of all the galaxy clusters are more or less affected by their local motions. If we assume that all the clusters contained in the most extreme sky region (Group A), have roughly the same peculiar velocity amplitude projected to our line-of-sight and towards the same direction (which is of course not to be expected), this projected peculiar velocity still needs to be $\sim 430\ \text{km/s}$ (moving away from us, or, in other words, all the redshifts of the Group A clusters need to be reduced by this velocity value) in order to explain the apparent anisotropy of Group A at a $2\sigma$ level. Finally, in all these cases, low-redshift samples such as ACC would be much more sensitive than high-redshift ones, such as XCS-DR1, which would not be significantly affected by such velocity amplitudes.  

\paragraph{Possibly different $N_{\text{HI}}$ structure.}

Keeping in mind that clusters with large $N_{\text{H,tot}}$ in their direction tend to have lower measured $L_X$ values, an uneven distribution of these clusters between Groups A and C could cause such a behavior. We should point out here that we do not claim a physical reason behind this; rather we state an observational result, as shown in the right panel of Fig. \ref{a-fit}. As displayed in Fig. \ref{NH} though, such an uneven distribution does not seem to be the case. The two groups have the same percentage of clusters with $N_{\text{HI}}>10^{21}\ \text{cm}^{-2}$ ($\sim 6$\%) and the rest of the clusters of the two groups have similar $N_{\text{HI}}$ values. Finally, the dependence of the derived $T$ on $N_{\text{H,tot}}$ will be tested in future work. In the case where  a multitemperature structure model needs to be considered, a significant Galactic absorption would unevenly affect the cold and hot thermal component, potentially biasing the measurement of $T$.

\paragraph{Cool-core clusters.}

A possible excess of cool-core clusters towards the Group A region could potentially result in higher $a$ values compared to sky regions with less cool-core clusters. Furthermore, a higher fraction of mergers towards a region could alter the obtained results for different regions as well. To this end, as a first test, we use the HIFLUGCS sample \citep{reiprich, hudson, mittal} which contains all the necessary information, including cool-core X-ray luminosities $L_{\text{cool}}$ for the 64 brightest galaxy clusters in the sky. For these clusters, 56 are also contained in ACC. Moreover, 46 clusters in total (72\% of the sample) have a cool-core, 16 of them lying within the Group A region (80\% of the total 20 clusters in Group A). Thus, the rest of the sample has 30 cool-core clusters out of a total of 46 members (68\%).  At the same time, 60\% (9 out of 15) of Group C clusters contain a cool-core. Group A contains 31\% of all the sample's clusters and 35\% of the cool-core ones. 

These small differences are not sufficient to explain the $\geq3\sigma$ deviations that we found for ACC. Moreover, these statistics are limited to a small fraction of the ACC clusters. Hence, they could be considered indicative for explaining these deviations only if the cool-core related results changed significantly compared to the general solution.
When we consider the bolometric $L_X$ for the entire sample, the deviation between Group A and the rest of the sample is surprisingly high (3.39$\sigma$, using again the Bootstrap method), while Group C is totally consistent ($<1\sigma$) with the rest of the sample for HIFLUGCS. Considering the cool-core-corrected luminosity $L=L_X-L_{\text{cool}}$ for the 46 clusters, the deviation between Group A and the rest persists with the same significance ($3.28\sigma$).  However, these deviation amplitudes are biased because of the quite different temperature distribution of Group A compared to the rest of the sky. Nonetheless, this strongly demonstrates that the obtained deviations are not the result of a possible bias added by an uneven distribution of cool-core clusters. 

\paragraph{Environmental effects, mergers, and superclusters.} \label{supercl-acc}

The properties of galaxy clusters, and particularly the $L_X-T$ scaling relation, may depend on the environment in which the clusters are formed. Studies have shown that disturbed clusters tend to be less luminous than undisturbed clusters, for the same X-ray temperatures  \citep{pratt,chon}. Therefore, a larger number of disturbed systems in the rest of the sky than in the Group A sky region could result in a lower normalization value of the $L_X-T$ relation, causing the tension between the two subsamples. On the other hand, a significant difference in the number of these systems between Group A and the rest would be necessary to account for $\sim 3\sigma$ anisotropies. If this were actually the case, another form of an anisotropy would occur, since there are no obvious reasons why such a difference should occur.

Disturbed clusters are expected to coincide with mergers and, in general, to be found in overdense regions, such as superclusters \citep{schuecker}. With the purpose of applying a first test, we now look at two subsamples from the 400d catalog, as given in \citet{400d}. There are 49 low-$z$ and 36 high-$z$ galaxy clusters, all observed with Chandra (while the 400d catalog was constructed by ROSAT \citep{rosat} archival data). The low-$z$ part is basically a subsample of the HIFLUGCS catalog, from where the 15 clusters with the lowest $z$ were excluded. All clusters come with an estimation of their dynamical state; for example, if they are relaxed or if they are merging systems. The vast majority of the low-$z$ clusters (45 or 92\%) are included in ACC as well.
We see that for the low-$z$ systems, Group A region contains $\sim 30\%$ of all the clusters and $\sim 25\%$ of all the mergers (covering 25\% of the whole sky). This shows that for this limited subsample, there are no obvious significant differences from the rest of the sky. From the 36 high-$z$ clusters, only 2 are lying within Group A, preventing us from comparing the number of mergers with the rest of the sample. 

Another test that could be applied is to look for superclusters in the sample and test the behavior of their cluster members, as well as their position in the sky. However, our ability to perform such a test is limited by the fact that ACC is mainly compiled by archival data from pointing observations of clusters, which were conducted for the purposes of different projects. This would add a significant bias in the process of finding superclusters, since some clusters could in fact be members of superclusters, but we would not identify them as such, since the other members of the supercluster would not have been included in the ACC catalog. This problem would be even more severe in the case where the linking length of the members of a supercluster were greater than the ASCA (GIS) field of view (FOV), namely $50\ \text{arcmin}$  \citep{tanaka}.  Considering that ACC is a low-$z$ sample, the vast majority of the supercluster members would be at distances greater than that (for the median $z=0.09$, the distance covered by the FOV of ASCA would be $\sim 5.5\  \text{Mpc}$). 

\subsection{XCS-DR1}

In order to verify this peculiar behavior of the specific sky regions as they have been identified for the ACC sample, we need to cross check our results with another, independent sample, namely XCS-DR1. 

\subsubsection{General solution and hemispheres}

\begin{figure*}[hbtp]
           \includegraphics[width=0.49\textwidth, height=6cm]{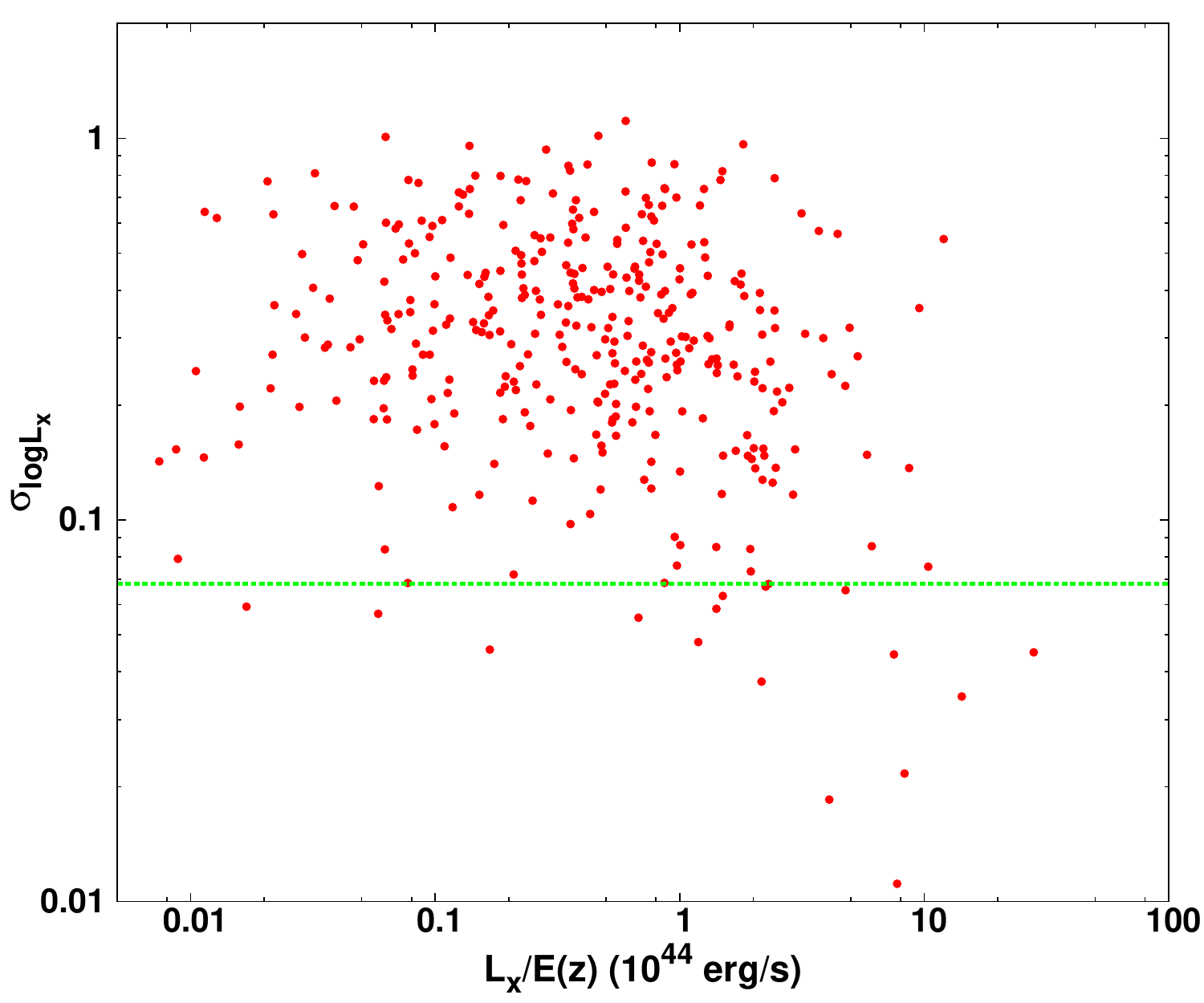}
            \includegraphics[width=0.49\textwidth, height=6cm]{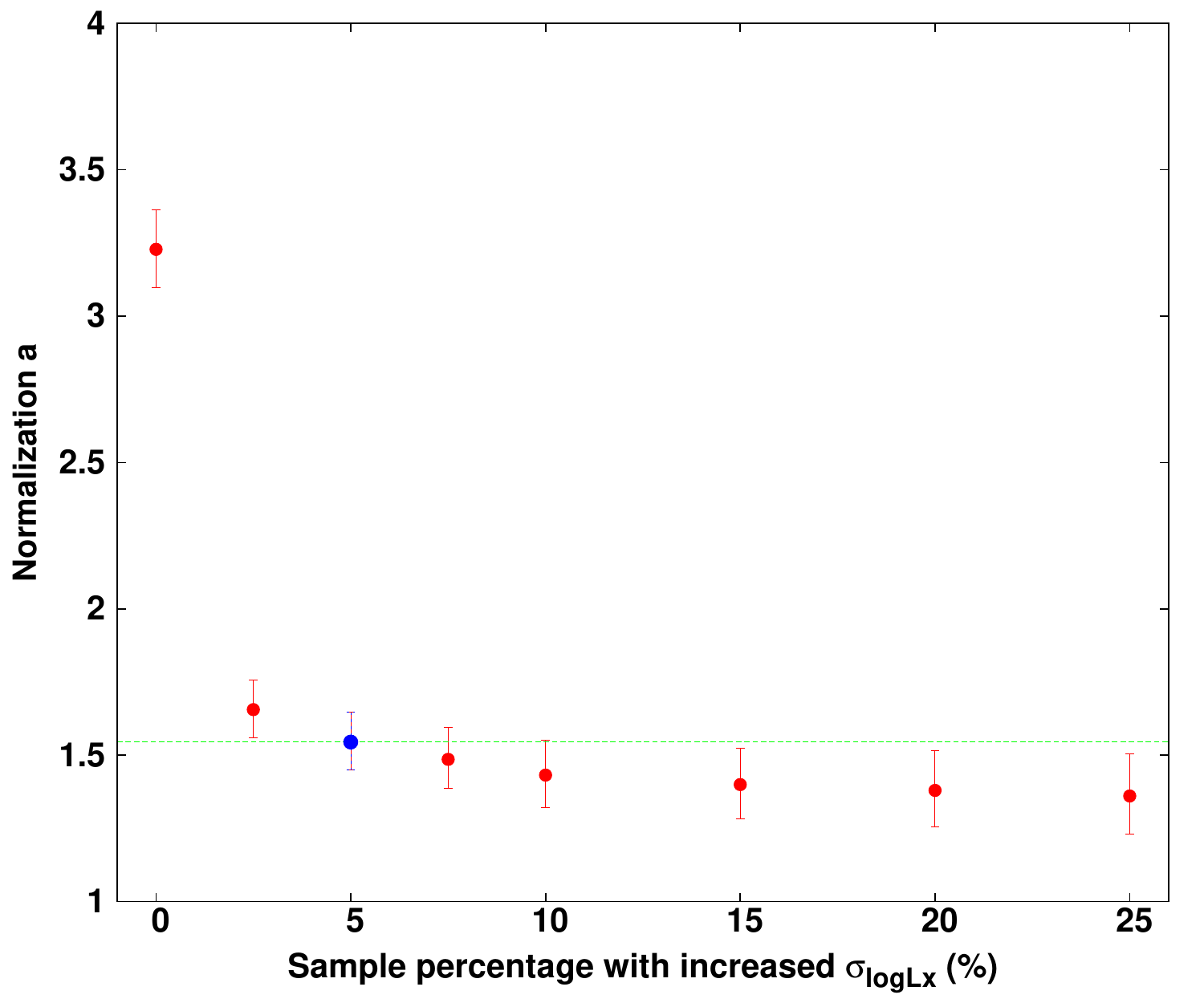}
          \caption{\textit{Left panel}: Logarithmic luminosity uncertainty $\sigma_{\log{Lx}}$ as a function of the X-ray bolometric luminosity $L_X$, together with the correction limit $\sigma_{\log{Lx}}=0.068$ (green line). \textit{Right panel}: The best-fit normalization value for the entire sample (with the slope fixed to its best-fit value) as a function of the sample percentage in which an extra $L_X$ uncertainty is added. The blue point and the green line represent the sample percentage that we use and the best-fit value we derive.}
        \label{fig6}
\end{figure*}

For this sample of galaxy clusters, the luminosity uncertainties $\sigma_{Lx}$ are more important than the temperature uncertainties $\sigma_T$.
Consequently, we now use Eq. (\ref{eq3}) for the fitting procedure. We consider the $\log$-uncertainties as described in Subsection \ref{fitting}, where $b_1=2.220$ is the best-fit value for the slope when we do not consider $\sigma_{\log{T}}$.

However, when we try to fit the model to the data, we notice the same behavior as in ACC. The normalization is fairly high ($a\sim 3$) and does not represent the sample well. This is again caused by the very low $\sigma_{\log{L}}$ and $\sigma_{\log{T}}$ values of some high$-L_X$ clusters $(L_X>10^{44}\ \text{erg/s})$. As a result, $\chi^2$ is once again overfitting these data, leading to a rather high normalization. To fix this, we follow a similar approach as in ACC, adding an extra "false" uncertainty to the 5\% of the sample with the lowest $\sigma_{\log{Lx,\text{final}}}$\footnote{${\sigma^2_{\log{Lx,\text{final}}}}={\sigma_{\log{Lx,\text{initial}}}^2}+(2.220\times \sigma_{\log{T}})^2$}. These clusters initially have $\sigma_{\log{Lx,\text{final}}}<0.068$  and we apply the "correction" such that they eventually have $\sigma_{\log{Lx,\text{final}}}=0.068$. These galaxy clusters also correspond  to $\sim 4-7\%$ of every subsample that we analyze later. The relative plot is shown in the left panel of Fig. \ref{fig6}. From the right panel of the same Figure, we can see the rapid change of the best-fit $a$ value once we add this extra $L_X$ uncertainty.

The final results for the entire XCS-DR1 sample and the four main Galactic hemispheres are shown in Table \ref{tab6}.
The noticeable features here are the much less steep slope compared to the ACC sample, the large uncertainties of the best-fit values, and the much lower $\chi^2_{min}/\text{d.o.f.}$ value, all because of the large uncertainties of the observations. Our results for the entire sample are in total agreement with those of \citet{hilton} (if one corrects for the redshift evolution that they fit), indicating the validity of our method.

Comparing the sky hemispheres, we do not find any deviations in their solutions (not larger than $\sim 1.5\sigma$).

\begin{table*}[hbtp]
\caption{\small{Best-fit values with their 3$\sigma$ credibility ranges for the four hemispheres, using Eq. (\ref{eq3}).}}
\label{tab6}
\begin{center}
\begin{tabular}{c  c  c  c}
\hline \hline

Hemisphere (No. of members, no. of corrected $\sigma_{\log{Lx}}$) & $a$ & $b$ & $\chi^2_{min}/\text{d.o.f.}$\\
\hline \hline

Northern (230, 11)  &\ $1.552^{+0.154}_{-0.140}$\ & \ $2.576\pm 0.119$ \ & $3.722$\\ 
Southern (134, 7) &\ $1.556^{+0.206}_{-0.182}$\ & \ $2.320\pm 0.214$\ & 2.766\\ \hline
Eastern (212, 9) &\ $1.667^{+0.182}_{-0.168}$\ & \ $2.386\pm 0.159$\ & 3.026\\ 
Western (152, 9) &\ $1.435^{+0.164}_{-0.147}$\ &\ $2.572\pm 0.138$\ & 3.752 \\ \hline
All (364, 18) &\ $1.545^{+0.118}_{-0.113}$\ &\ $2.512\pm 0.104$\ &  3.388\\
 \hline \hline 
\end{tabular}
\end{center}
\end{table*}

\subsubsection{Different sky solid angles}

\begin{figure*}[h!]
          \includegraphics[width=0.49\textwidth, height=6cm]{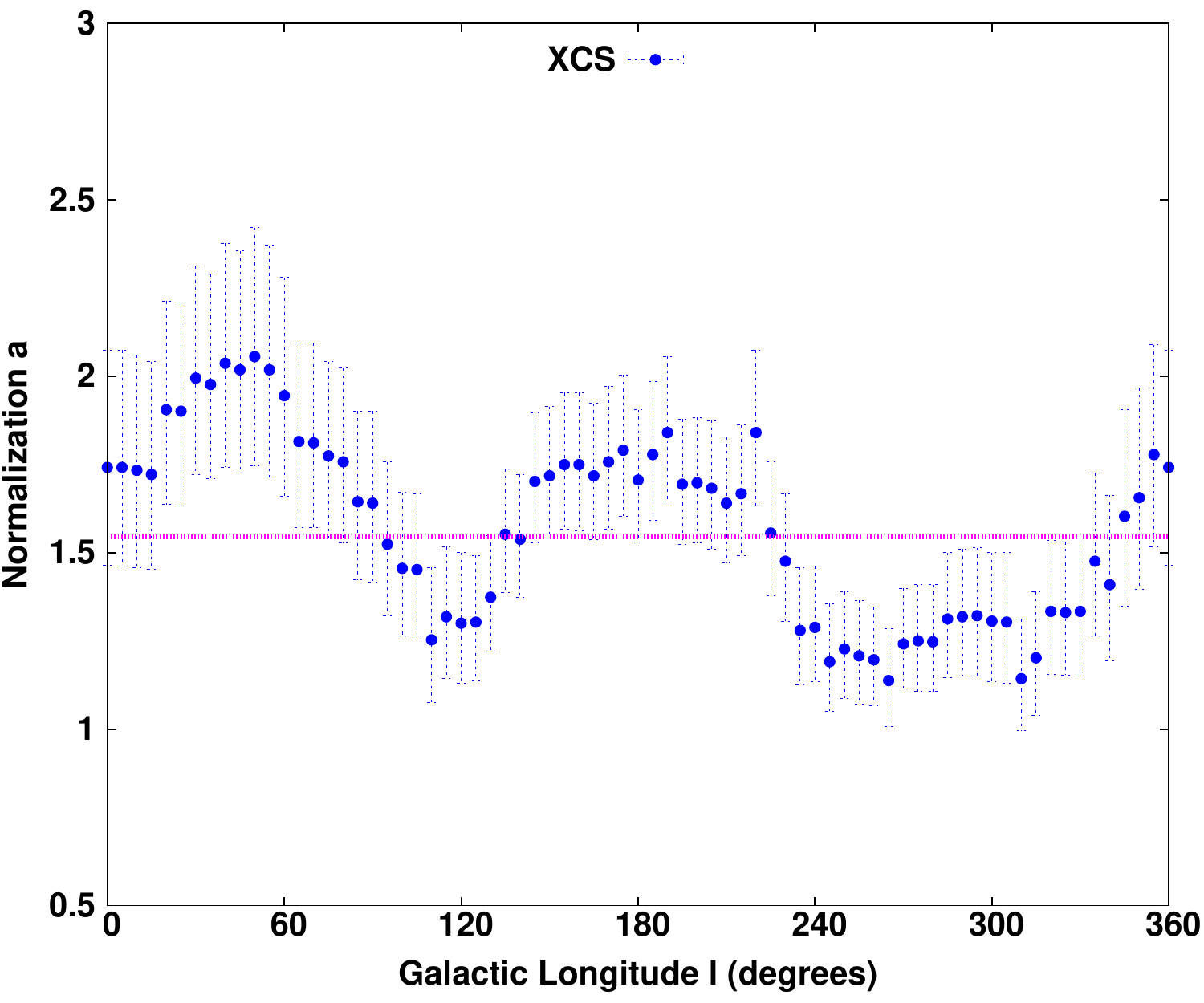}
             \includegraphics[width=0.49\textwidth, height=6cm]{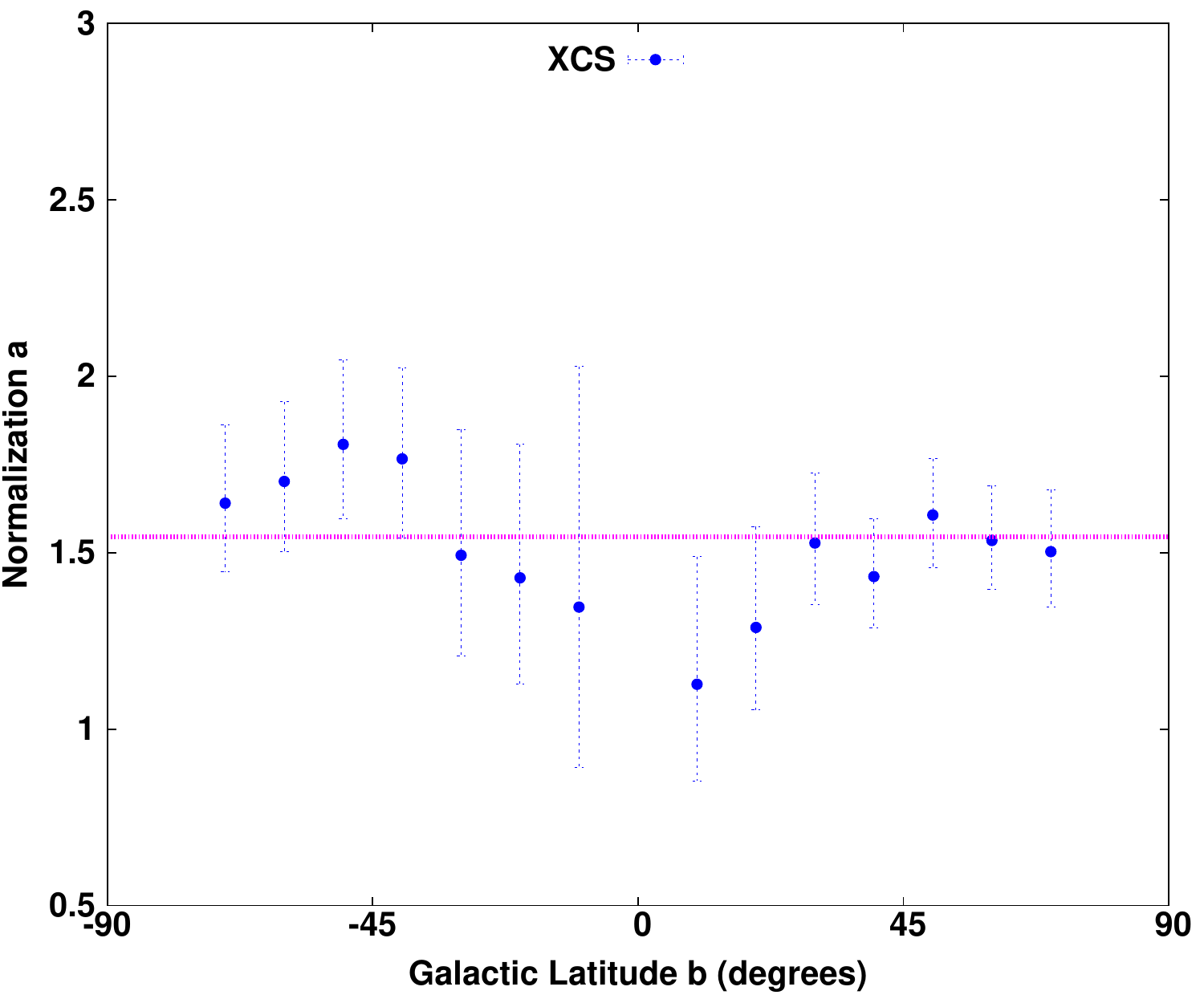}
            \includegraphics[width=0.49\textwidth, height=6cm]{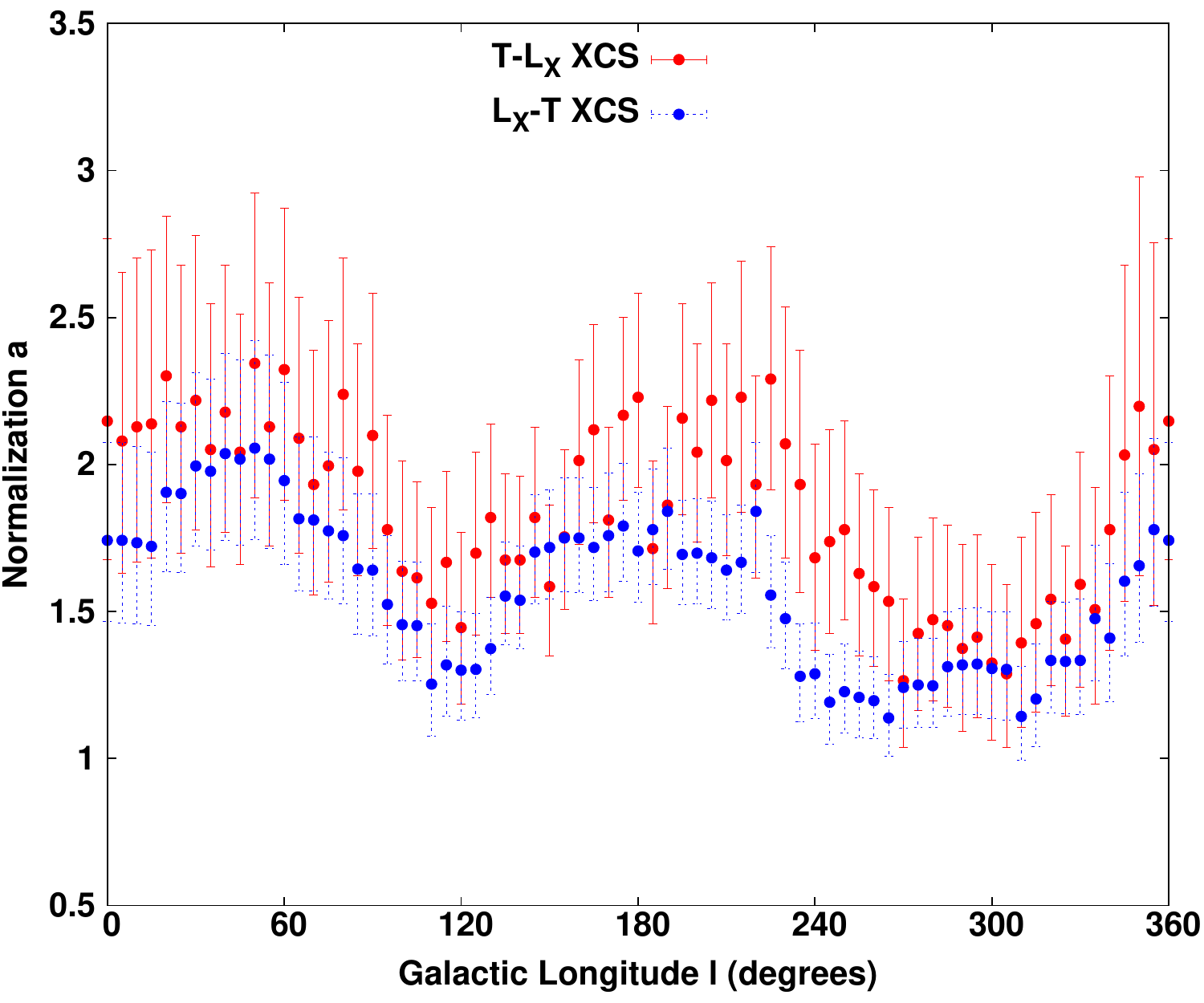}
             \includegraphics[width=0.49\textwidth, height=6cm]{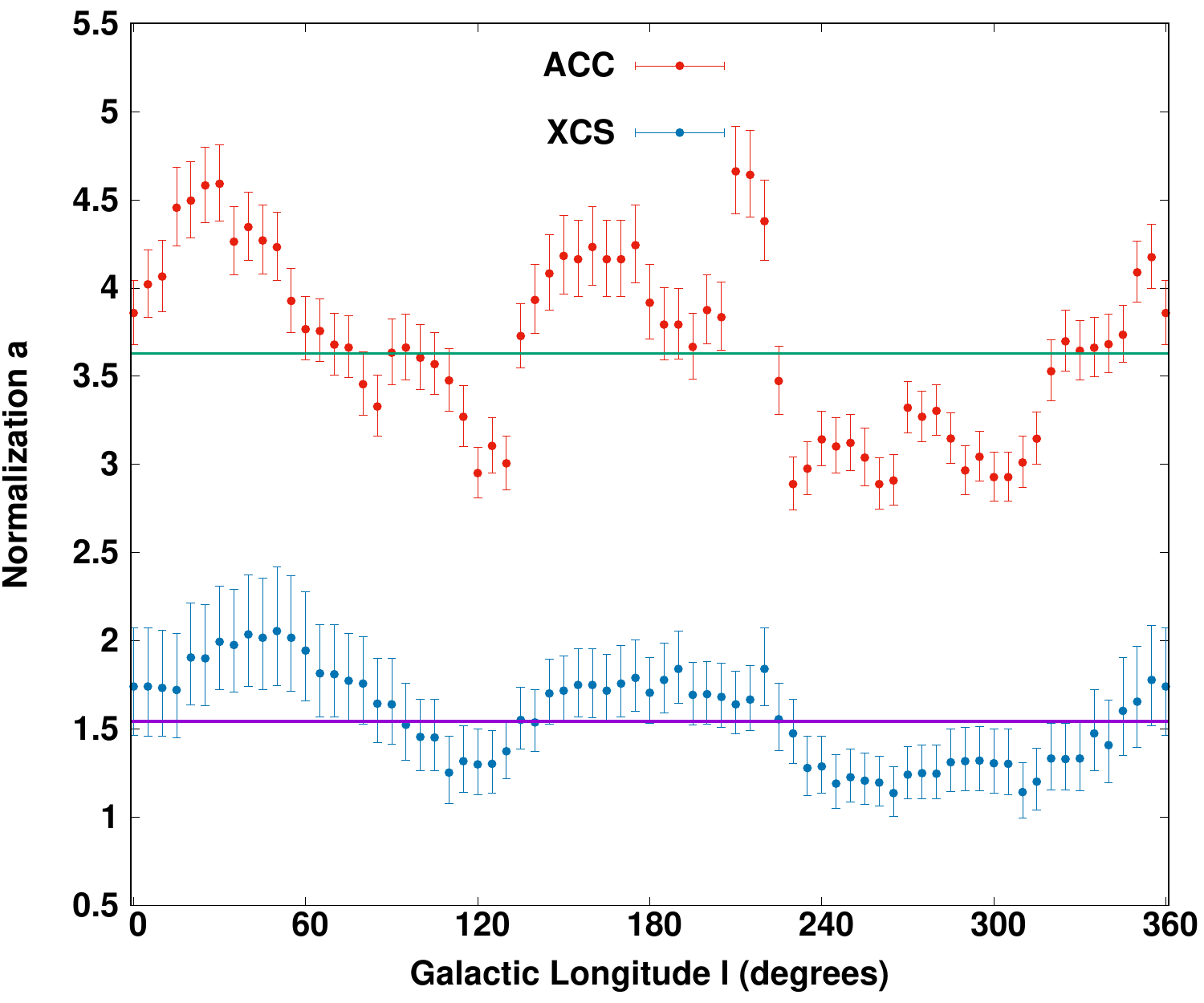}
            \caption{Best-fit value of the normalization with its $3\sigma$ uncertainty for every sky region of XCS-DR1: \textit{Top left:} with $\Delta l=90^{\circ}, \ \Delta b=180^{\circ}$ as a function of its central Galactic longitude, with the best-fit normalization value for the whole sample (purple). \textit{Top right:} with $\Delta l=360^{\circ}, \ \Delta b=40^{\circ}$ as a function of its central Galactic latitude. \textit{Bottom left:} for the $L_X-T$ (blue) and $T-L_X$ (red) fitting. \textit{Bottom right:} for XCS-DR1 (blue) and ACC (red) with their best-fit values for the whole sample (green line for ACC and purple line for XCS-DR1). Given the bin widths, the data points are obviously strongly correlated.}
        \label{XCS-a-fit}
\end{figure*}

We repeat the steps we applied during the ACC analysis, scanning the sky as described in Subsection  \ref{identif_anis}, obtaining the normalization. During this, we fix the slope to $b=2.388$. The results are displayed in Fig. \ref{XCS-a-fit}. 

The clusters of the XCS-DR1 sample are not evenly distributed throughout the Galactic longitudes, since all the regions which are entirely included within $l\in (90^{\circ},240^{\circ})$ contain more than 140 members each. At the same time, the rest of the regions could contain as few as 52 clusters. Groups A and B contain 78 and 83 objects, respectively. 

As shown in the top right panel of Fig. \ref{XCS-a-fit}, in the case of XCS-DR1, $a$ does not appear to have large differences for different Galactic latitude regions. The only region that could imply an inconsistent behavior is the one with $b\in(-70^{\circ},-30^{\circ})$, which ends up having a $2.01\sigma$ tension with the rest of the sample. At the same time, all the other regions have a $<1\sigma$ tension with the rest of the sample.
XCS-DR1 does not contain any clusters with $|b|<20^{\circ}$. 

The most characteristic feature of Fig. \ref{XCS-a-fit} (bottom right panel) is the very similar fluctuation of $a$ with the Galactic longitude as in ACC. The normalization value follows approximately the same fluctuation pattern as in ACC, with the highest normalization being found in the sky region within $l\in (-5^{\circ},95^{\circ})$, which we will also refer to as Group A for the XCS-DR1 sample. The corresponding region for ACC was lying within $l\in (-20^{\circ},75^{\circ})$ while the lowest values were found within $l\in (75^{\circ},175^{\circ})$ and $l\in (255^{\circ},340^{\circ})$. We remind the reader that XCS-DR1 shares only three common clusters with ACC, while only one is within the Group A regions, not causing a significant effect. Therefore, the two samples do not have an obvious reason to have the same behavior.

For XCS-DR1, the lowest $a$ value belongs to the sky region within $l\in (265^{\circ},355^{\circ})$, almost identically with ACC. However, its statistical significance, which we obtain after the performed analysis, is not sufficiently high ($1.99\sigma$), due to the few cluster members it contains (60). Therefore, practically the lowest $a$ belongs to the region within $l\in (200^{\circ},310^{\circ})$, which we will refer to as Group B. The region within $l\in (95^{\circ},155^{\circ})$ (which we call Group C for XCS-DR1), which is a subpart of the Group C region as defined for the ACC sample, has also a low normalization, with a somewhat lower statistical significance than Group B. Originally, the region with the lowest $a$ at that part of the sky, was $l\in (65^{\circ},155^{\circ})$ ($\Delta l=90^{\circ}$). However, since we do not wish this region to overlap with Group A, we set the low limit to $l=95^{\circ}$, which in fact, gives a stronger deviation. Nevertheless, the main low-$a$ sky part of XCS-DR1, remains Group B.

\begin{table*}[hbtp]
\caption{\small{Galactic longitude $l$ limits for Groups A, B, and C as they are defined for ACC and XCS-DR1. All groups have $b\in [-90^{\circ},90^{\circ}]$. In parenthesis, the number of clusters they contain and the deviation of each Group compared to the rest of the sample are displayed, as obtained by Bootstrap. The $\pm$ symbols indicate if each Group's normalization is greater or smaller than the rest of the sample.}}
\label{tab7}
\begin{center}
\begin{tabular}{c  c  c}
\hline \hline

Sky region & ACC & XCS-DR1\\
\hline \hline

Group A  & \ $l\in [-20^{\circ},75^{\circ}]\ \ (75, +2.65\sigma)$ \ & $l\in [-5^{\circ},95^{\circ}]\ \ (78, +3.12\sigma)$\\ 
Group B & \ $l\in [215^{\circ},310^{\circ}]\ \ (58, -1.86\sigma)$\ & $l\in [200^{\circ},310^{\circ}]\ \ (83, -2.67\sigma)$\\
Group C &\ $l\in [75^{\circ},175^{\circ}]\ \ (82, -2.06\sigma)$ \ & $l\in [95^{\circ},155^{\circ}] \ \ (81, -2.33\sigma)$\\ 
 \hline 
\end{tabular}
\end{center}
\end{table*}

The large uncertainties of this sample lead to a $\chi^2_{red}$ of $\sim 3$, significantly lower than the one of ACC. However, this again does not allow us to directly infer the deviations between subsamples only by the 3$\sigma$ uncertainties of their derived best-fit results (using the $\Delta\chi^2$ limits), since this would require $\chi^2_{red}\sim 1$. Moreover, different scatter for different subsamples would cause $\chi^2_{red}$ to vary, affecting the best-fit values of uncertainties derived from the $\Delta\chi^2$ limits. This could sometimes result in similar (or even smaller) 3$\sigma$ uncertainties for the best-fit values of subsamples with significantly less clusters than others. Hence, we again apply the Bootstrap and Jackknife methods to ensure that there are no outliers causing this behavior and to properly assess the statistical significance of any deviation.

\begin{figure}[hbtp]
          \includegraphics[width=0.49\textwidth, height=6cm]{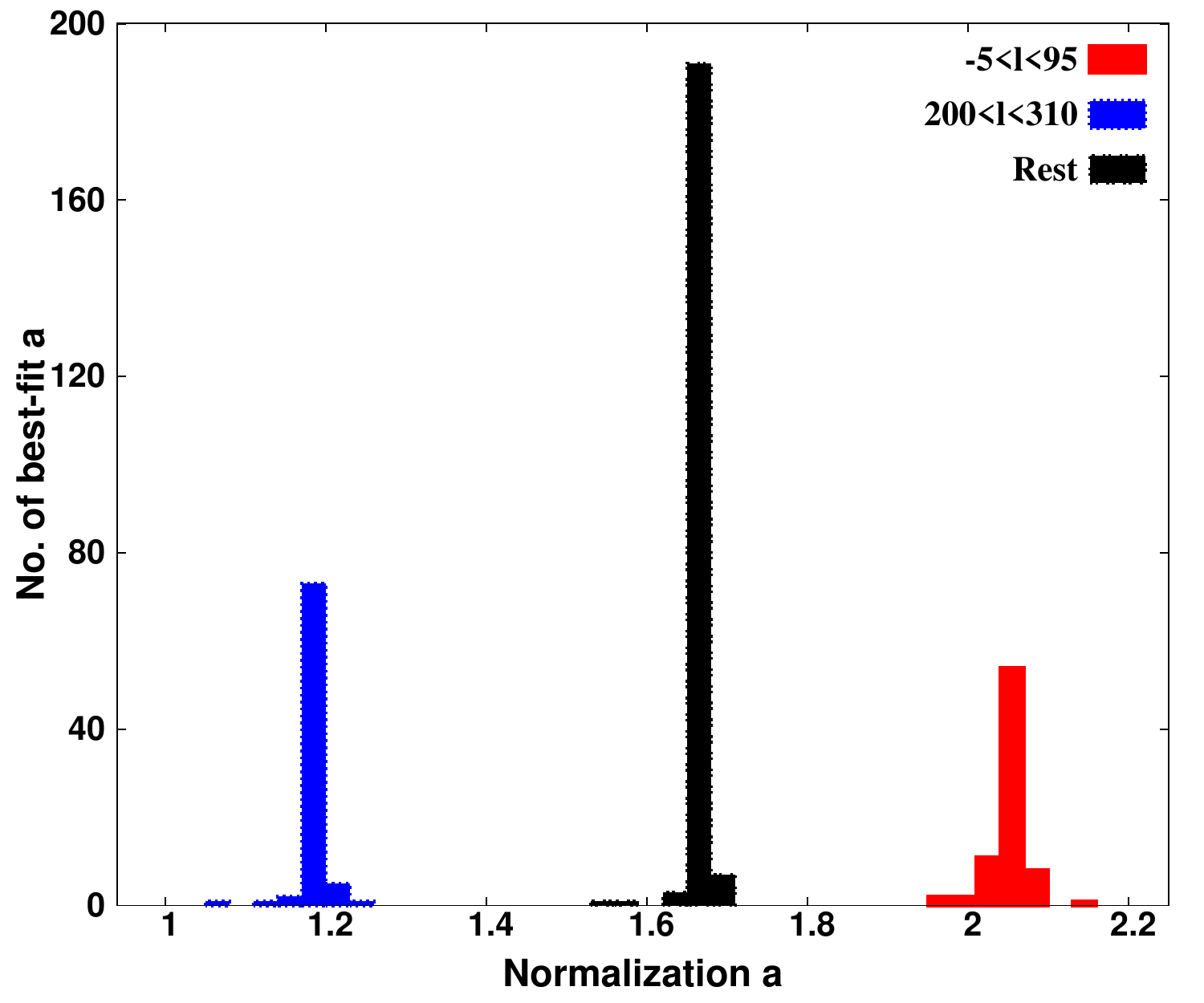}

        \caption{Distribution of the best-fit values of the normalization as obtained by the Jackknife resampling method, for Group A (red), Group B (blue), and the rest of the sample (black).}
        \label{jack-2}
\end{figure}

Due to the larger number of clusters XCS-DR1 contains and the nearly ten times larger uncertainties than ACC, the results are not affected by single data, as shown in Fig. \ref{jack-2}.

For Group A, we obtain  $a=2.051^{+0.337}_{-0.289}$, which intensely deviates by $3.12\sigma$ from the rest of the sample. If we exclude the galaxy cluster with the most strong effect on the Group A behavior, cluster XMMXCS J2339.8-1213, the deviation shifts to $2.73\sigma$. 

Group B has $a=1.183^{+0.141}_{-0.126}$ and a discrepancy of $2.67\sigma$ with the rest of the sample. Excluding XMMXCS J0056.0-3732, the discrepancy moves to $2.44\sigma$. 

If we consider the rest of the sample subtracting both Groups A and B, the deviation of Group A is now $1.69\sigma$, while for Group B it is $2.22\sigma$. Considering a typical size for the random subsamples of 80 clusters while we apply the Bootstrap method, we find that the tension between Group A and Group B is at $3.89\sigma$, for which the statistical significance is impressively large (again bearing in mind the small sample size). Even if we exclude the two above-mentioned galaxy clusters, the deviation is no less than $3.34\sigma$. 

In the bottom-left panel of Fig. \ref{XCS-a-fit} we also display the results for the $T-L_X$ fitting, which gives us a slope of $b=3.066\pm 0.137$. In this case, the sensitivity to individual clusters is larger, making the dispersion and the uncertainties of the derived best-fit values also larger. Due to these uncertainties, all the deviations we refer to between Group A, Group B, and the rest of the sample drop to $45\%-65\%$ for the $T-L_X$ fitting. However, since the luminosity uncertainties are much larger than the temperature uncertainties, the $L_X-T$ fitting is more appropriate here. 

\subsubsection{Free slope}

Now we consider the slope as a free parameter, in order to compare the $a-b$ solution space between the two Groups and the rest of the sample. In Table \ref{tab7} the results of the fit are displayed, where we see that the slope of Group A also differs to a large degree. 

\begin{table*}[hbtp]
\caption{\small{Best-fit values with their 3$\sigma$ credibility ranges for Groups A and B and the rest of the sample, using Eq. (\ref{eq3}).}}
\label{tab7}
\begin{center}
\begin{tabular}{c  c  c  c}
\hline \hline

Subsample (No. of members, no. of corrected $\sigma_{\log{Lx}}$) & $a$ & $b$ & $\chi^2_{min}/\text{d.o.f.}$\\
\hline \hline

Group A (78, 4) &\ $1.742^{+0.333}_{-0.280}$\ & \ $2.032\pm 0.254$ \ & $2.109$\\ 
All without Group A (286, 14) &\ $1.489^{+0.129}_{-0.118}$\ & \ $2.606\pm 0.113$\ & 3.485\\
\hline
Group B (83, 6) &\ $1.294^{+0.178}_{-0.157}$\ & \ $2.714\pm 0.160$ \ & $ 3.665$\\ 
All without Groups A and B (203, 8) &\ $1.644^{+0.188}_{-0.169}$\ & \ $2.438\pm 0.164$\ & 3.166\\
 \hline 
\end{tabular}
\end{center}
\end{table*}

\begin{figure*}[hbtp]
             \includegraphics[width=0.49\textwidth, height=6cm]{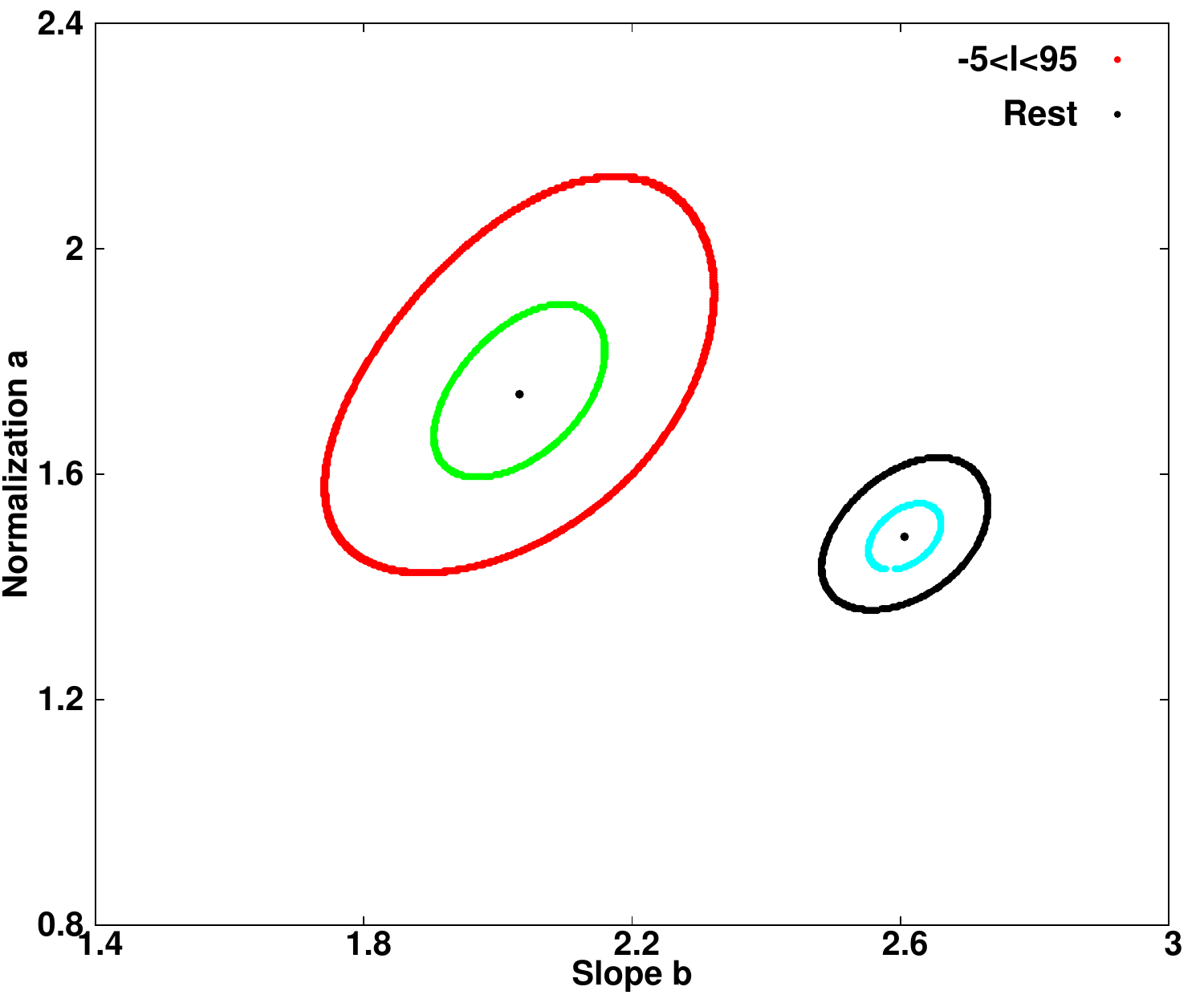}
           \includegraphics[width=0.49\textwidth, height=6cm]{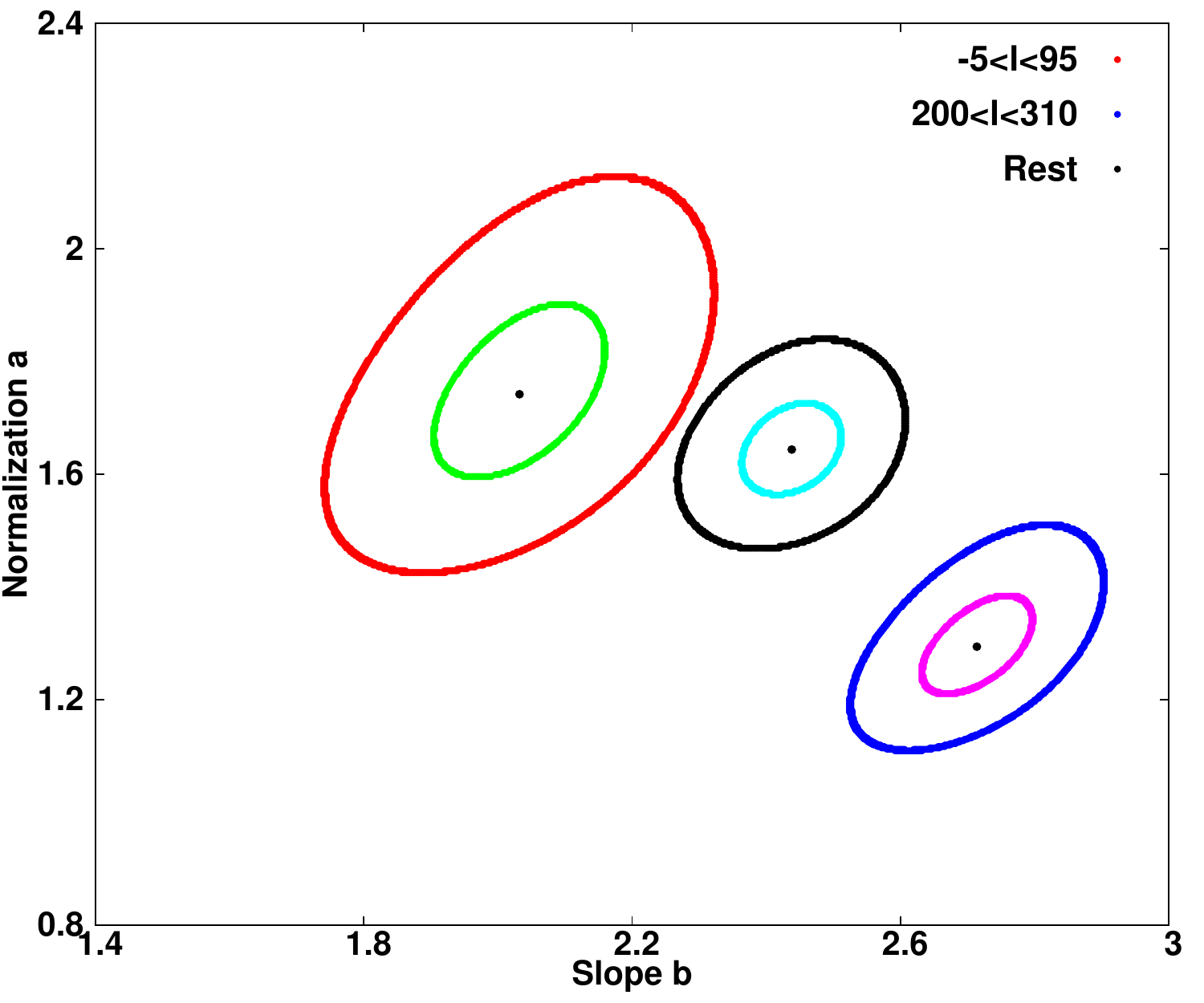}
           \caption{1$\sigma$ (green, light blue and purple)  and 3$\sigma$  (red, black and blue) contour plots, where the intrinsic scatter is not considered. \textit{Left:} For Group A and the rest of the sample, including Group B. \textit{Right}: Group A, Group B, and the rest of the sample, excluding Group B.}
        \label{xcs-plane}
\end{figure*}

As shown in Fig. \ref{xcs-plane}, despite the large uncertainties in $\log{L_X}$ (in the order of $\sim 100\%$ adding $\sigma_{\log{T}}$ as explained), Group A does not share any common $3\sigma$ solution space with Group B or the rest of the sample in general. Additionally, Groups A and B share very limited 3$\sigma$ common solution space with the rest of the sample when both are excluded from it.

\begin{figure}[hbtp]
          \includegraphics[width=0.49\textwidth, height=6cm]{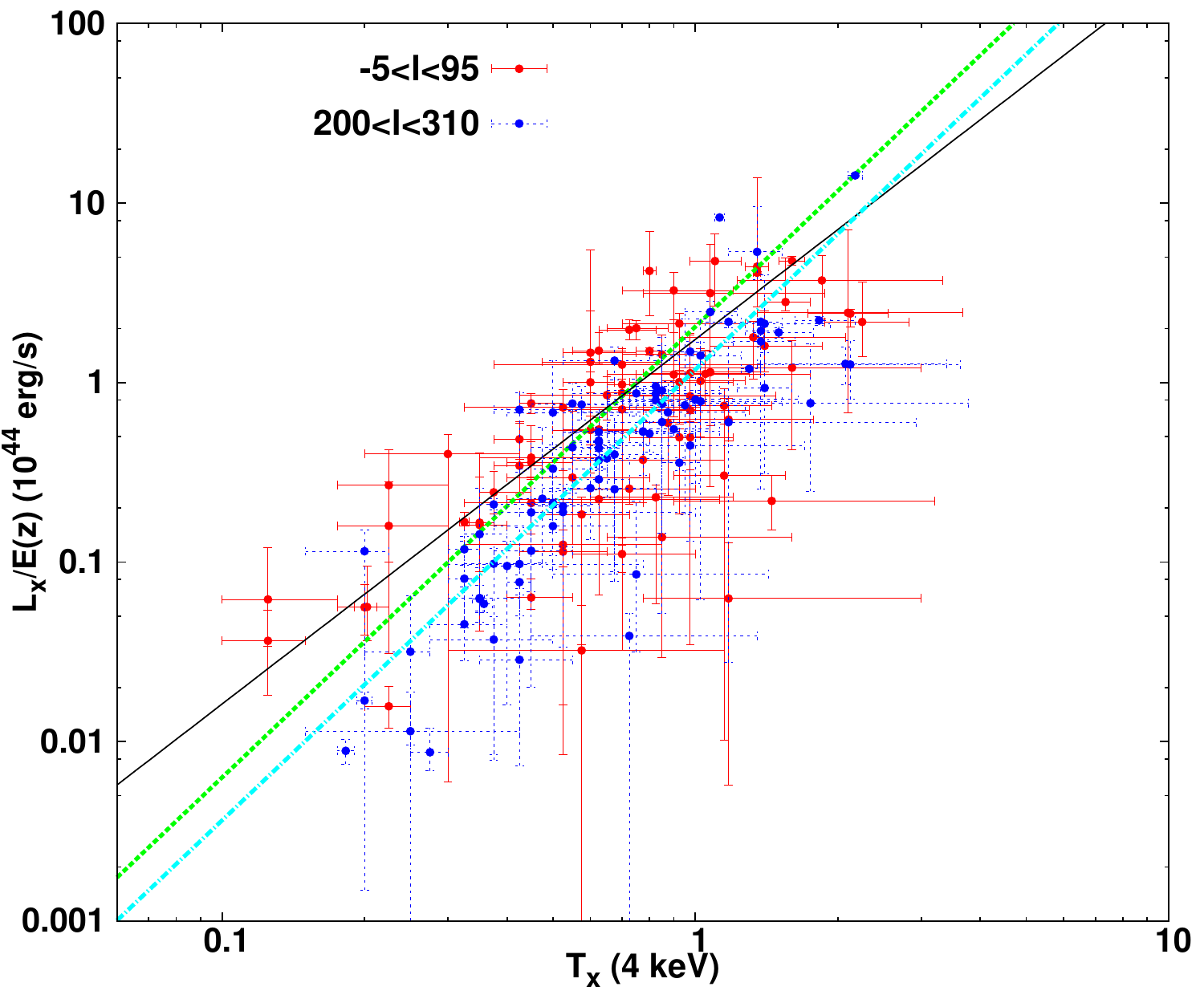}

        \caption{Bolometric luminosity $L_X$ as a function of temperature $T$ for Group A (red) and Group B (blue). The best-fit functions are also displayed for Group A [$a=2.051,\  b=2.512$ (fixed) with green and $a=1.742,\  b=2.032$ (free) with black] and for Group B [$a=1.183,\  b=2.512$ (fixed) with light blue].}
        \label{L-T-XCS}
\end{figure}

In Fig. \ref{L-T-XCS}, the $L_X-T$ plot of the two Groups is displayed.

\subsubsection{Possible causes} \label{causes-XCS}

\paragraph{Distributions of $T,\ z$ , and $\sigma_{\log{Lx}}$.}

As we did for ACC, we test if the apparent anisotropy is caused by different distributions of the temperatures, redshifts, or uncertainties of the Groups or by a specific temperature or redshift range. From Table \ref{tab3}, we see that the tension between the two Groups and the rest of the sample is consistent for all the redshift ranges. However, Group B and the rest of the sample have the same solution for high$-z$ clusters. Furthermore, this is also the case for Group A and the rest of the sample, but for high$-T$ clusters. Therefore, the deviation seems to be due to the low$-T$ clusters.

\begin{table*}[hbtp]
\caption{\small{Best-fit normalization values with their 3$\sigma$ credibility ranges for Group A, Group B, and the rest of the sample of the XCS-DR1 sample, for low and high temperature and redshift ranges.}}
\label{tab3}
\begin{center}
\begin{tabular}{ c  c  c  c  c  c}
\hline \hline

Subsample (No. of members)  & All & $T\leq 3$ keV & $T>3$ keV & $z\leq 0.35$ & $z>0.35$\\
\hline \hline

Group A (78, 4) &\ $2.051^{+0.337}_{-0.289}$\ & \ $2.582^{+0.617}_{-0.498}$ \ & $1.614^{+0.395}_{-0.317}$ & $2.023^{+0.392}_{-0.333}$ & $2.128^{+0.723}_{-0.540}$\\ 
Group B (83, 6) & $1.183^{+0.141}_{-0.126}$&\ $1.074^{+0.185}_{-0.156}$\ & \ $1.312^{+0.229}_{-0.198}$ \ & $1.049^{+0.153}_{-0.131}$ & $1.567^{+0.365}_{-0.293}$ \\ \hline
Rest (203, 8) &\ $1.667^{+0.165}_{-0.150}$\ & \ $1.710^{+0.285}_{-0.241}$\ & $1.641^{+0.209}_{-0.185}$ & $1.819^{+0.246}_{-0.216}$ & $1.489^{+0.229}_{-0.198}$\\
 \hline 
\end{tabular}
\end{center}
\end{table*}

As argued above, such a behavior could also be caused by a different redshift, temperature, or observable uncertainties distribution of the subsamples. XCS-DR1 is a homogeneous sample though, with the same selection functions towards all sky directions.

\begin{figure*}[hbtp]
              \includegraphics[width=0.32\textwidth, height=5cm]{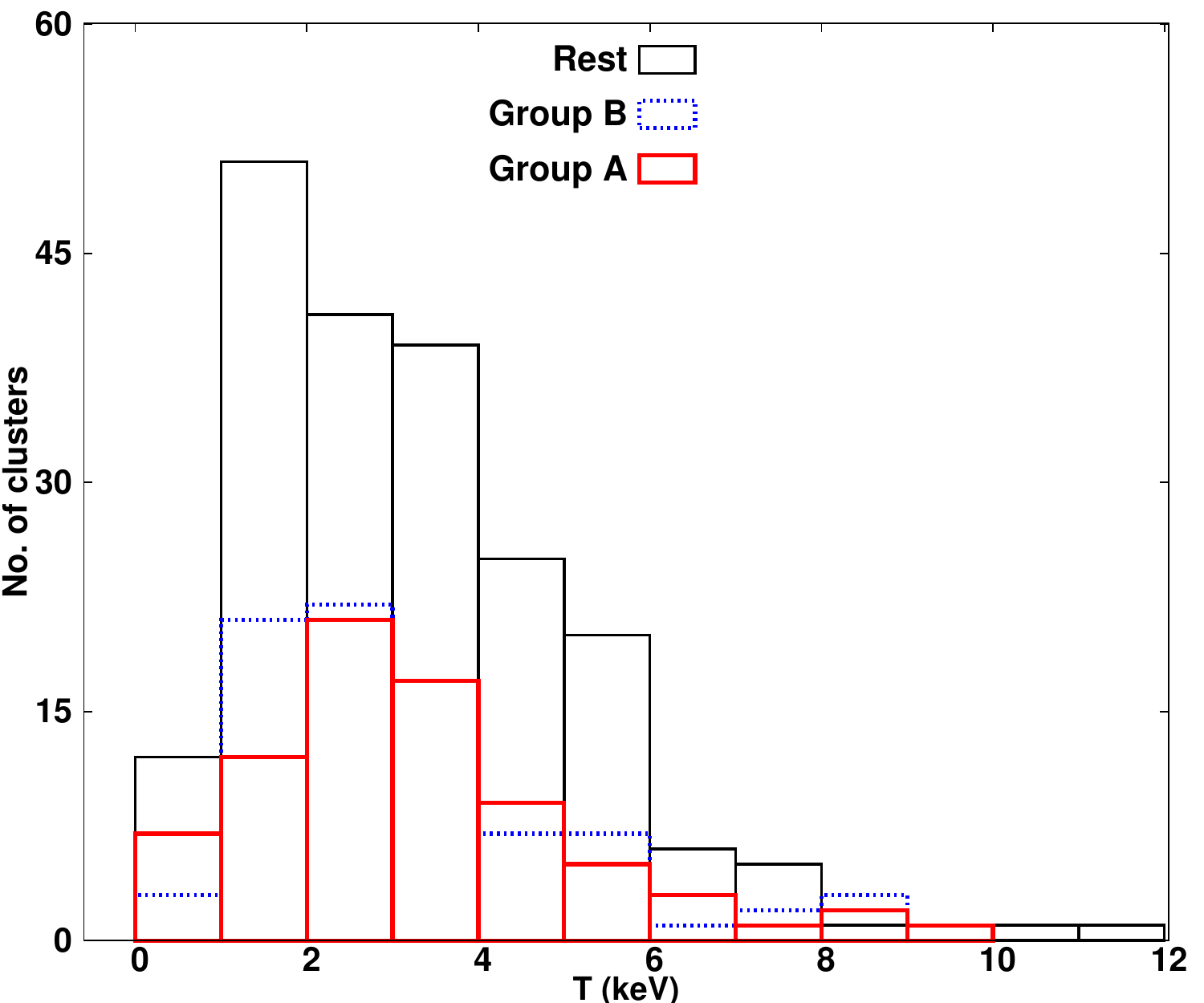}
            \includegraphics[width=0.32\textwidth, height=5cm]{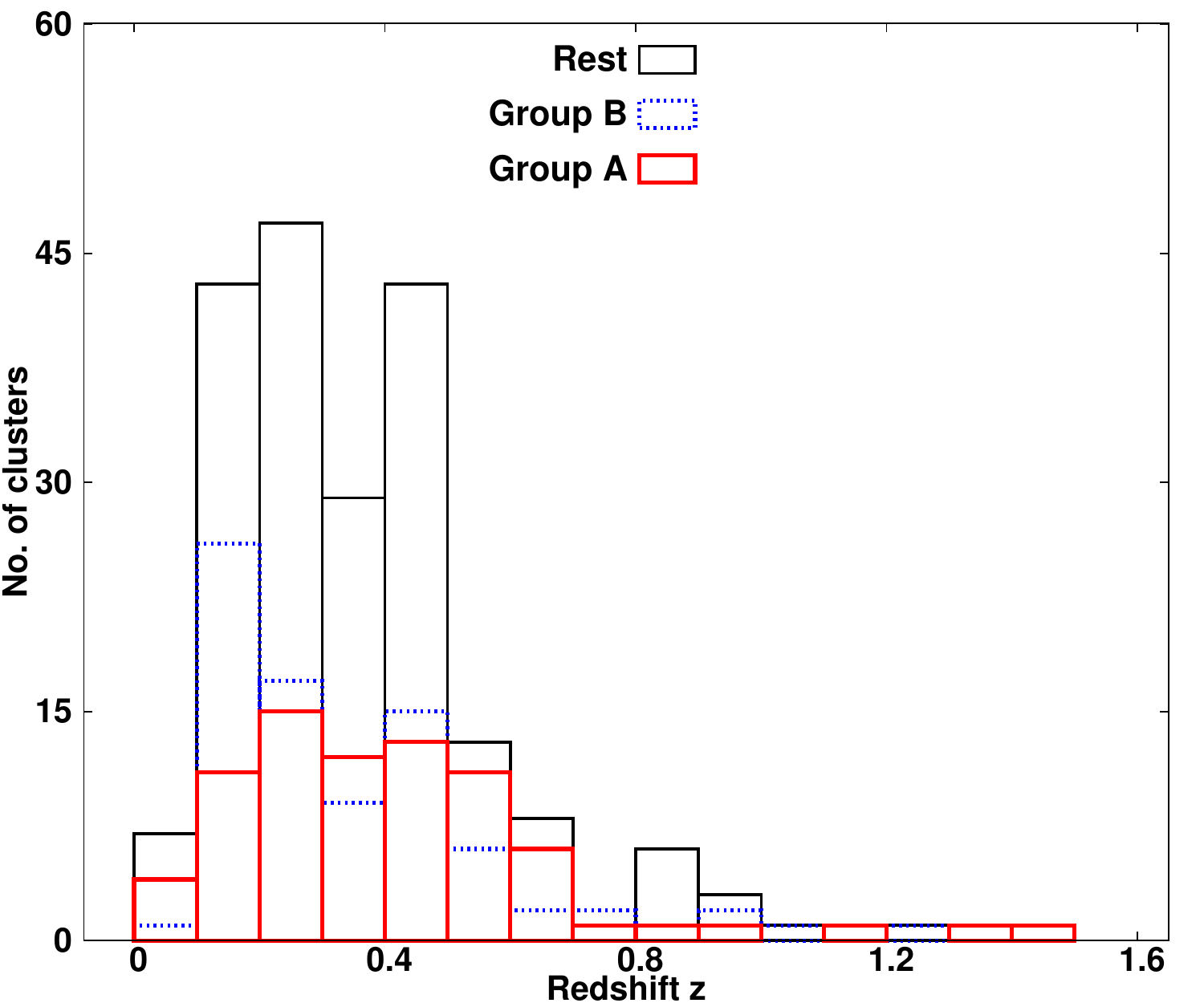}
            \includegraphics[width=0.32\textwidth, height=5cm]{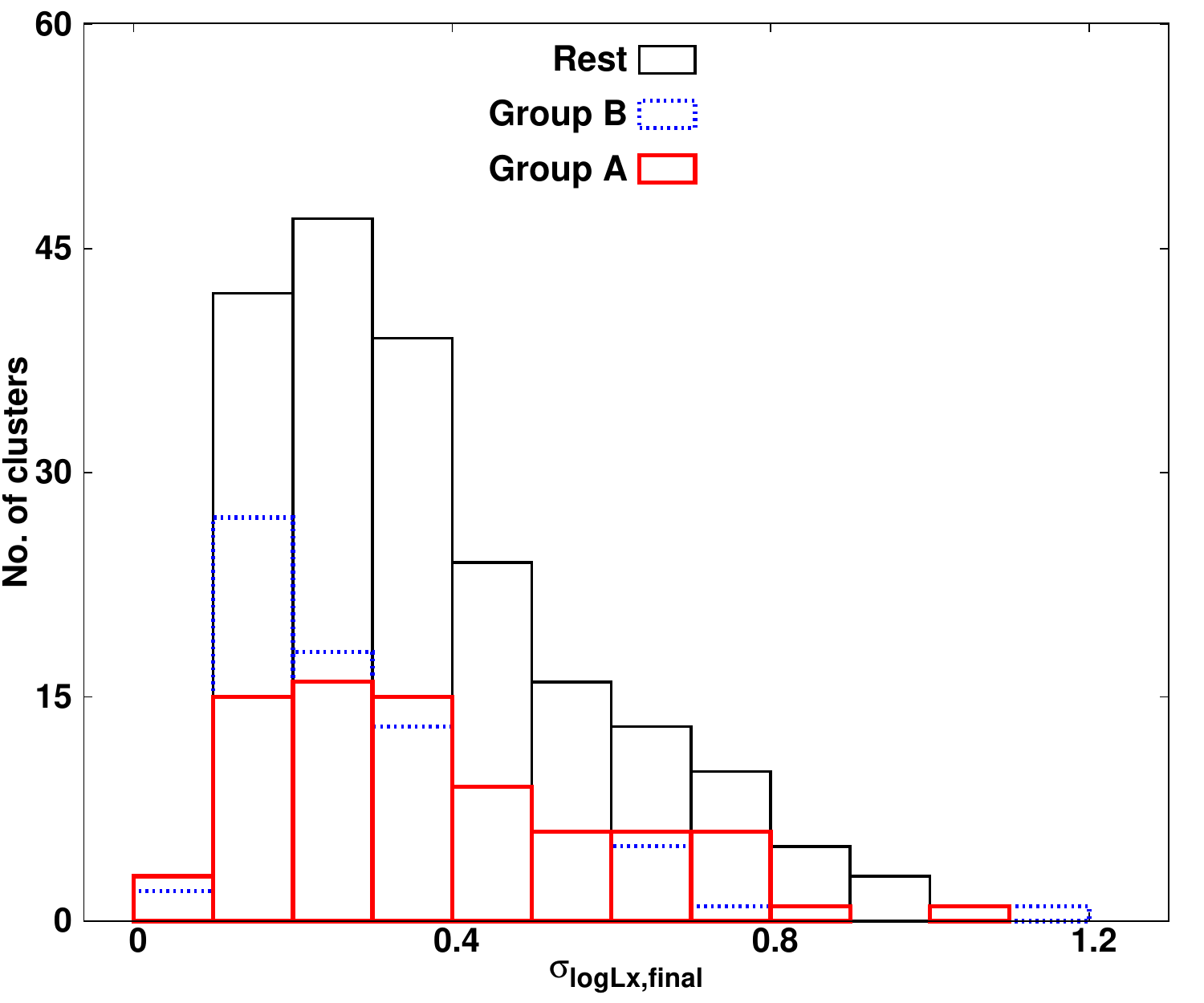}
            \caption{Distribution of the temperature (left), redshift (center) and final luminosity uncertainties (right) of the galaxy clusters contained in Group A (red), Group B (blue), and the rest of the sample (black).}
        \label{distrib}
\end{figure*}

As illustrated in Fig. \ref{distrib}, and as it was expected, this does not seem to be the case. All the distributions of Group A and the rest of the sample without Group B, are quite similar. On the contrary, Group B has more low$-z$ galaxy clusters than Group A, but this is not the cause of the deviation between the two groups, since the latter is relatively consistent throughout all the redshift ranges, as shown in Table \ref{tab3}. 

Thus, the reason for the strong deviation of mainly the Group A region and secondarily the Group B and C regions, that exists in both samples, remains to be found.

\paragraph{Redshift correction to the CMB frame and peculiar velocities.}

XCS-DR1 mostly contains clusters at high redshifts and thus their measured redshift values are not practically affected by the bulk motion of our Galaxy or the peculiar velocities of the clusters. With the purpose of demonstrating this analytically, we repeat the tests we performed for ACC. In order for Group A and the rest of the sample to be consistent at a 2$\sigma$ level, the bulk velocity of our Galaxy would need to be $u_{bulk}\gtrsim 11000$ km/s towards  $(l,b)\sim (270^{\circ}, 35^{\circ})$ or $u_{bulk}\gtrsim 3800$ km/s (the minimum possible) towards $(l,b)\sim (210^{\circ}, 0^{\circ})$. Moreover, if we again make the same assumption (as we also did for ACC), where we consider that all the Group A clusters have the same projected peculiar velocity to our line-of-sight, this velocity needs to be $\sim 2800$ km/s in order to decrease the tension at 2$\sigma$ (instead of $3.12\sigma$ that it was originally). Obviously all these peculiar velocity values are unreasonable in almost all cosmological models, meaning that the explanation for the obtained deviation  does not seem to be based on the wrongly determined cosmological redshifts due to peculiar motions of the objects.

\paragraph{Environmental effects and superclusters}

As explained before for the ACC sample, identifying clusters that belong in rich environments such as superclusters (and are possibly disturbed), and studying their $L_X-T$ behavior, could indicate if these systems could cause the apparent deviations we observe between Group A and the rest of the sky.  Since the XCS-DR1 sample is composed of clusters that were serendipitously detected on the archival observations of XMM-Newton data bases, there is a higher probability of identifying multiple members of the same supercluster, than in ACC. Of course the selection bias which was described in Section \ref{supercl-acc} will again affect our results, but to a lesser degree. This bias is eliminated in the case in which the supercluster members are separated by less than the FOV of XMM-Newton, namely $\sim 30\ \text{arcmin}$.  Moreover, one has to consider that XCS-DR1 is a high-$z$ sample, and thus the FOV covers greater proper distances at the cluster redshifts ($\sim 10$ Mpc for the median $z=0.315$); however still not enough for the vast majority of superclusters.

To this end, we perform a first, simple test to try to identify superclusters and investigate if they have any effect on the apparent anisotropy we obtain from the data. Using the coordinates and the redshifts of the clusters as well as a specific linking length $R$ of our choice, we identify all cluster pairs separated by distances $\le R$. Then, starting from a pair, we find all the clusters which are connected to that pair, iteratively adding clusters until no more can be found at distances $\le R$. This is considered an isolated supercluster structure. Repeating these steps for all the initial pairs, we finally obtain all the clusters that belong to such structures. Of course, the number of such structures heavily depends on our choice of $R$.  

Starting with $R=20\ \text{Mpc}$, we only find 10 superstructures, containing 21 clusters in total. It is rather obvious that the number of clusters is too small to properly evaluate their behavior, since the statistical uncertainties are quite large. Thus, we increase the linking length to $R=50\ \text{Mpc}$, this time obtaining 22 superstructures containing 48 galaxy clusters (most of them in pairs).  For these clusters, we obtain $a=1.603 \pm 0.311$ and $b=2.969 \pm 0.222$, and if we fix the slope to $b=2.512$, we then obtain $a=1.274 \pm 0.234$. This clearly shows that supercluster members are underluminous compared to field clusters. However, their sky distribution is not quite uniform, since 10 out of the 48 clusters are found in the XXL North part of the sky, a $25\ \text{deg}^2$ area fully observed by XMM-Newton with high exposure times, for the needs of the XXL survey \citep{xxl}. Moreover, out of these 48 clusters, only 5 lie within Group A, which is $6.5\%$ of all the clusters of this group. At the same time, for the rest of the sky, $15\%$ of all the clusters belong to superclusters. If we exclude the clusters contained in the XXL north region, which clearly adds a bias since $65\%$ of its clusters belong to superclusters, then for the rest of the sky we find a fraction of $12.5\%$. There is a slight difference between Group A and the rest, but the statistics are too limited to derive a conclusive result. Nevertheless, the clusters that appear to belong in superclusters, seem to be significantly more luminous for Group A (5 clusters with $a=2.001\pm 1.143$) compared to the rest (43 clusters with $a=1.197\pm 0.230$). If we entirely exclude all the 48 clusters that seem to belong in superclusters and redo the analysis, the deviation of Group A decreases from $3.12\sigma$ to $2.50\sigma$, which is still statistically significant. Finally, Groups B and C have once more the most underluminous clusters that belong to superstructures.

We now increase the linking length of the superclusters to $R=70\ \text{Mpc}$, finding  29 superclusters, containing in total 63 galaxy clusters. For these, we obtain $a=1.517 \pm 0.240$ and $b=2.804 \pm 0.212$. Fixing the slope to $b=2.512$, we obtain $a=1.346 \pm 0.195$, so, again supercluster members appear underluminous compared to field clusters. Out of these 63 clusters, 11 belong to Group A (14\% of all its clusters) and 52 to the rest of the sky (18\%). Once again, the XXL North region contains 18 supercluster members, that is, 85\% of its clusters. Excluding once again this region, the rest of the sky has a supercluster-members fraction of  13\%. For this $R$, we see that the Group A region has practically the same fraction of supercluster members as the rest of the sky does, excluding or not the XXL North region. It is noteworthy that when we fit the 11 clusters of Group A, we obtain $a=2.285\pm 0.889$, which is consistent with the general behavior of the Group A region, while for the rest of the sky we obtain $a=1.188\pm 0.217$. For these clusters, Groups B and C have $a=0.953\pm 0.285$ (13 clusters) and $a=0.831^{+1.010}_{-0.453}$ (11 clusters), respectively. All in all, even the clusters that belong to rich environments such as superclusters, and are possibly disturbed, seem to share the same behavior trends for all the different sky patches, returning the similar differences in the normalization value as the ones we obtained during the initial analysis.  If we once again exclude these 63 clusters from our analysis, the deviation between Group A  and the rest of the sky shifts to $2.32\sigma$. This small drop of the statistical significance is mostly due to the smaller size of Group A this time (67 clusters instead of 78) rather than the more consistent $a$ values between the two subsamples. The interpretation from all these could be that the cause of the deviation of Group A compared to the rest of the sky, is also affecting the clusters located within rich environments, rather than be caused by them. This could also be seen by the similar fraction of such clusters between Group A and the rest. 

\subsection{Cosmological constraints}

During our analysis up to now, we assumed fixed cosmological parameters towards all the directions in the sky in order to derive the normalization and slope of the $L_X-T$ relation. If this time we assume that the normalization and slope should be the same for every subsample of galaxy clusters (using the best-fit value for the whole sample), we can express any deviation that occurs in terms of the cosmological parameters, that is, $\Omega_{\text{m}}$ and $H_0$. These enter through the conversion of the X-ray flux to luminosity (via the luminosity distance) while $\Omega_{\text{m}}$ also enters through the redshift correction factor $E(z)$, as in Eq. (\ref{eq1}). We should note that we do not care about the actual values of the the cosmological parameters, since they will be biased due to the $a, \ b$ values which were determined using a specific cosmological model. Furthermore, we do not know precisely whether the actual intrinsic evolution is indeed self-similar. What we aim to compare are simply the occurring deviations between the $\Omega_{\text{m}}$ and $H_0$ values for different subsamples. 
Generally, these parameters are strongly correlated with the normalization. 

\subsubsection{$\Omega_{\text{m}}$ fitting}

Repeating all the steps firstly with respect to the $\Omega_{\text{m}}$ fitting, we find the $\Omega_{\text{m}}$ deviations to be roughly of the same amplitude as for the normalization. In fact, for the XCS-DR1 sample the $\Omega_{\text{m}}$ deviations are slightly larger, while for ACC, the Bootstrap results are not Gaussian and the exact statistical deviation cannot be determined.

Towards Group A region, we obtain a much higher $\Omega_{\text{m}}$ value than for the rest of the sky. This could indicate the need for smaller cosmological distances for a fixed redshift towards this sky region, leading to lower $L_X$ values of the clusters, and eventually alleviating the tension in the normalization of $L_X-T$ between this region and the rest of the sky.

\begin{figure*}[hbtp]
           \includegraphics[width=0.49\textwidth, height=6cm]{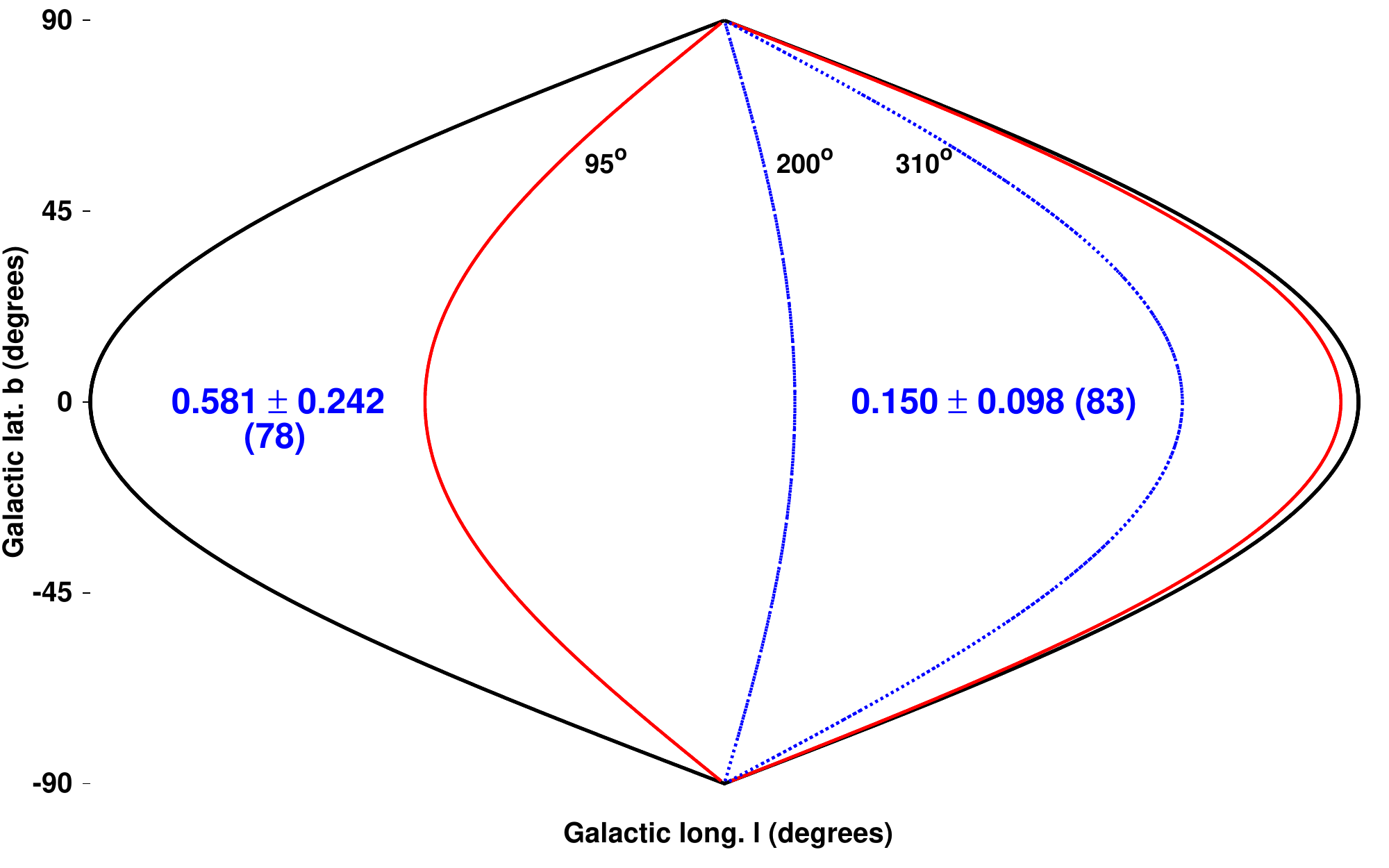}
            \includegraphics[width=0.49\textwidth, height=6cm]{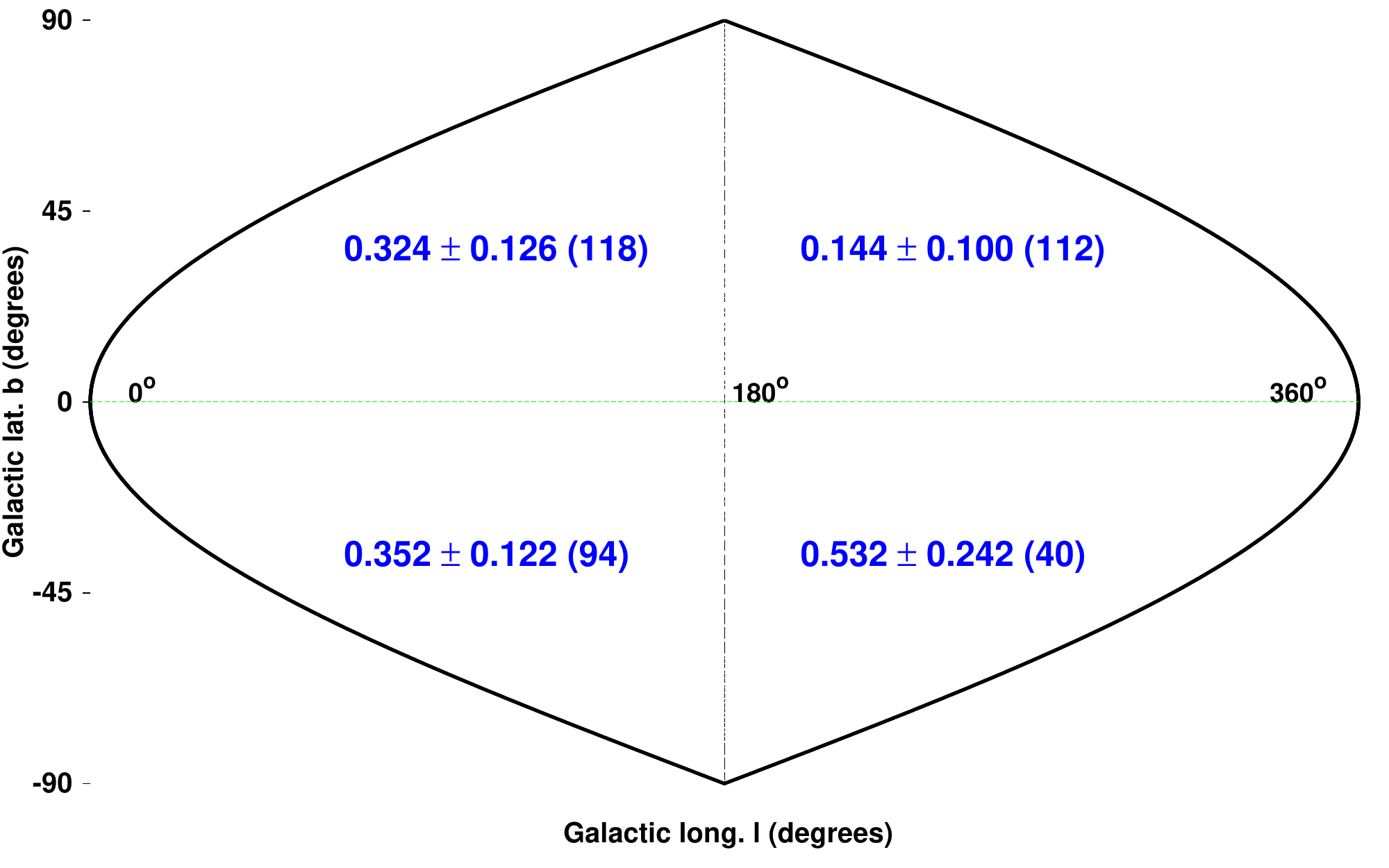}
          \caption{Best-fit value of $\Omega_{\text{m}}$ for XCS-DR1, with its $3\sigma$ uncertainty and the number of clusters in each region. Different uncertainty magnitudes for similar numbers of clusters are caused by the different scatter of the $L_X-T$ relation}. \textit{Left panel}: For Groups A and B. The rest of the sky has $\Omega_{\text{m}}=0.318\pm 0.071$.  \textit{Right panel}: For the four "usual" quarters of the Galactic sky.
        \label{cosmo}
\end{figure*}

For instance, for the entire XCS-DR1 sample, we obtain $\Omega_{\text{m}}=0.300\pm 0.061$, while for Group A this is $\Omega_{\text{m}}=0.581\pm 0.242$ and for the rest of the sample this is $\Omega_{\text{m}}=0.264\pm 0.064$, giving us a $3.02\sigma$ tension. More specifically, Group B has $\Omega_{\text{m}}=0.150\pm 0.098$ ($2.28\sigma$ from the rest of the sample) and Group C gives $\Omega_{\text{m}}=0.076^{+0.141}_{-0.071}$ ($2.05\sigma$), not being as statistically significant as Group B. The Galactic coordinates map of these regions together with the results are shown in the left panel of Fig. \ref{cosmo}. Additionally, in the right panel of Fig. \ref{cosmo}, the best-fit values for $\Omega_{\text{m}}$ are displayed for different patches of the sky. Interestingly, the lowest value in the right panel occurs from the sky region in which the axis for the maximum acceleration (minimum $\Omega_{\text{m}}$) of the Hubble expansion has been detected by several authors and methods (see references in Section \ref{intro}), usually with low to mild statistical significance. Moreover, the "warm" end of the CMB dipole roughly coincides with the center of this region. The higher $\Omega_{\text{m}}$ occurs from the same $l$ in the southern Galactic hemisphere, but it contains very few clusters in order to draw a trustworthy result.

\begin{figure}[hbtp]
           \includegraphics[width=0.49\textwidth, height=6cm]{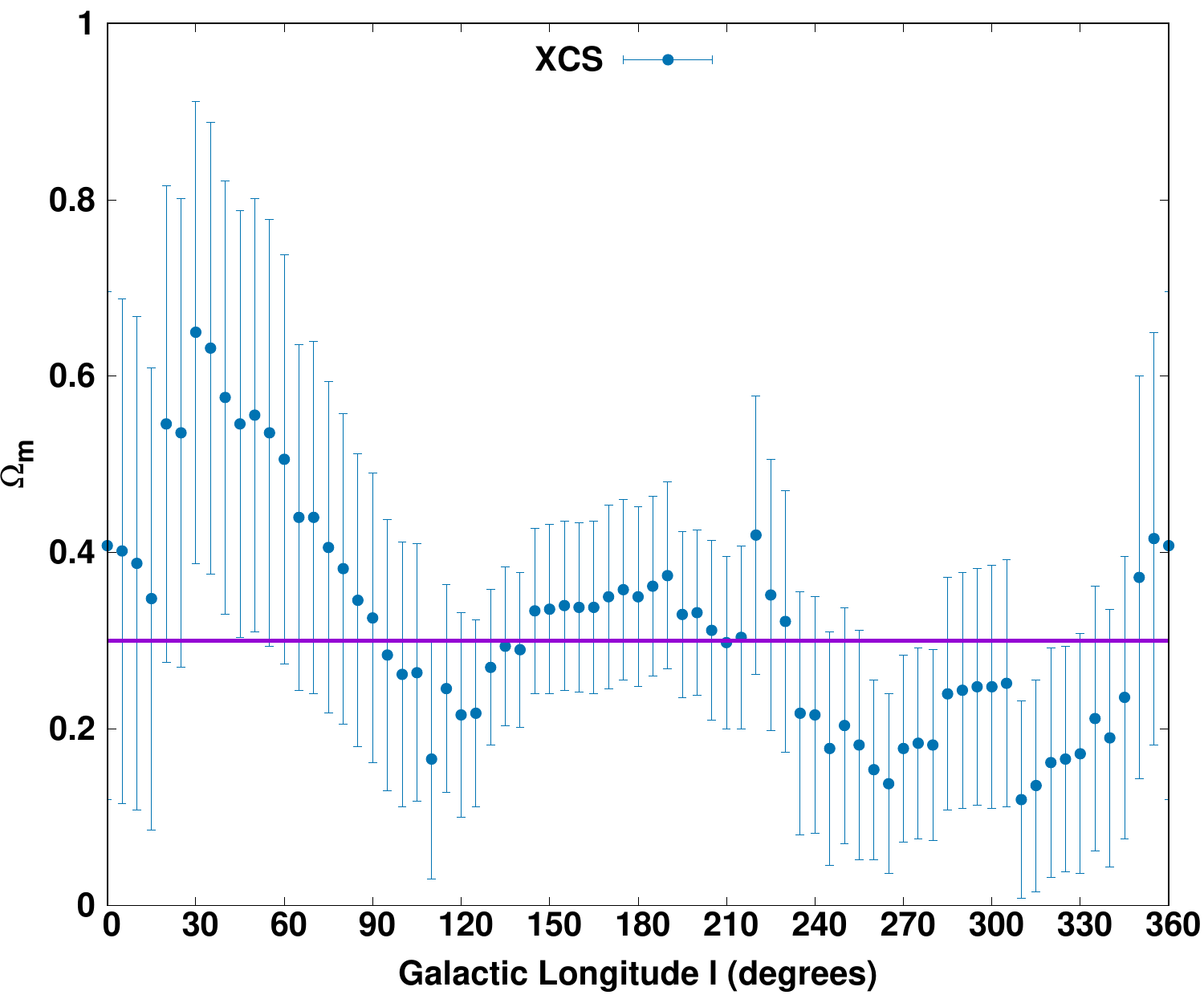}
          \caption{Best-fit value of $\Omega_{\text{m}}$ for every sky region of XCS-DR1 with $\Delta l=90^{\circ}, \ \Delta b=180^{\circ}$ as a function of its central Galactic longitude.}
        \label{matter-fit}
\end{figure}

Repeating the "scanning" of the sky for $\Omega_{\text{m}}$ as we previously did for $a$, we obtain the results shown in Fig. \ref{matter-fit}. From this, we see that the largest deviation for $\Omega_{\text{m}}$ does not occur for what we defined as Group A for XCS-DR1, but is found within $l\in [-15^{\circ},75^{\circ}]$, which is almost the same sky region with Group A as defined for ACC. The $\Omega_{\text{m}}$ value for this sky patch is $\Omega_{\text{m}}=0.650\pm 0.262$ and it appears to have a $3.53\sigma$ tension with the rest of the sample (which has $\Omega_{\text{m}}=0.258\pm 0.067$), according to the Bootstrap results. 

Moreover, the lowest $\Omega_{\text{m}}=0.120\pm 0.120$ appears for the sky region within $l\in [265^{\circ},345^{\circ}]$ (where we "cut" the region $l\in (345^{\circ},355^{\circ}]$ so it does not overlap with the highest-$a$ region). However, it has a low statistical significance ($1.44\sigma$) due to its limited number of clusters (50). Thus, the lowest-$\Omega_{\text{m}}$ region turns out to be $l\in [75^{\circ},155^{\circ}]$ (where we excluded the first 10$^{\circ}$ to avoid overlapping regions), which has $\Omega_{\text{m}}=0.074_{-0.074}^{+0.120}$ with a deviation of $3.03\sigma$ from the rest of the sample. Therefore, the general sky region of Group A remains the one with the most intense inconsistency for the $\Omega_{\text{m}}$ fitting as well. 

XCS-DR1 is more suitable than ACC for the $\Omega_{\text{m}}$ fitting, due to its  larger $z$ distribution. For clusters with low$-z$, such as those contained in ACC, the $\Omega_{\text{m}}$ value cannot be easily constrained, since it does not change significantly the $L_X$ or the $E(z)$ factor of a cluster.

With this in mind, for the entire ACC sample, we obtain a value of $\Omega_{\text{m}}=0.412\pm 0.091$, while for Group A we get $\Omega_{\text{m}}=1.112^{+0.220}_{-0.268}$. At the same time, for the rest of the sample we obtain $\Omega_{\text{m}}=0.176^{+0.096}_{-0.084}$. Group A's value is greater than all of the results derived by Bootstrap. Group C returns a value of $\Omega_{\text{m}}=0^{+0.072}_{-0.000}$ while Group B gives $\Omega_{\text{m}}=0.252^{+0.181}_{-0.160}$.

\subsubsection{$H_0$ fitting}

If one wishes to study the apparent deviations based on a purely kinematic approach, the most obvious option would be to only fit $H_0$, which directly probes the expansion rate of the local Universe. Since the effect of $H_0$ on $L_X$ through the luminosity distance does not depend on the redshift, both ACC and XCS-DR1 samples can be used to give us valuable results.  

\begin{figure*}[hbtp]
           \includegraphics[width=0.49\textwidth, height=6cm]{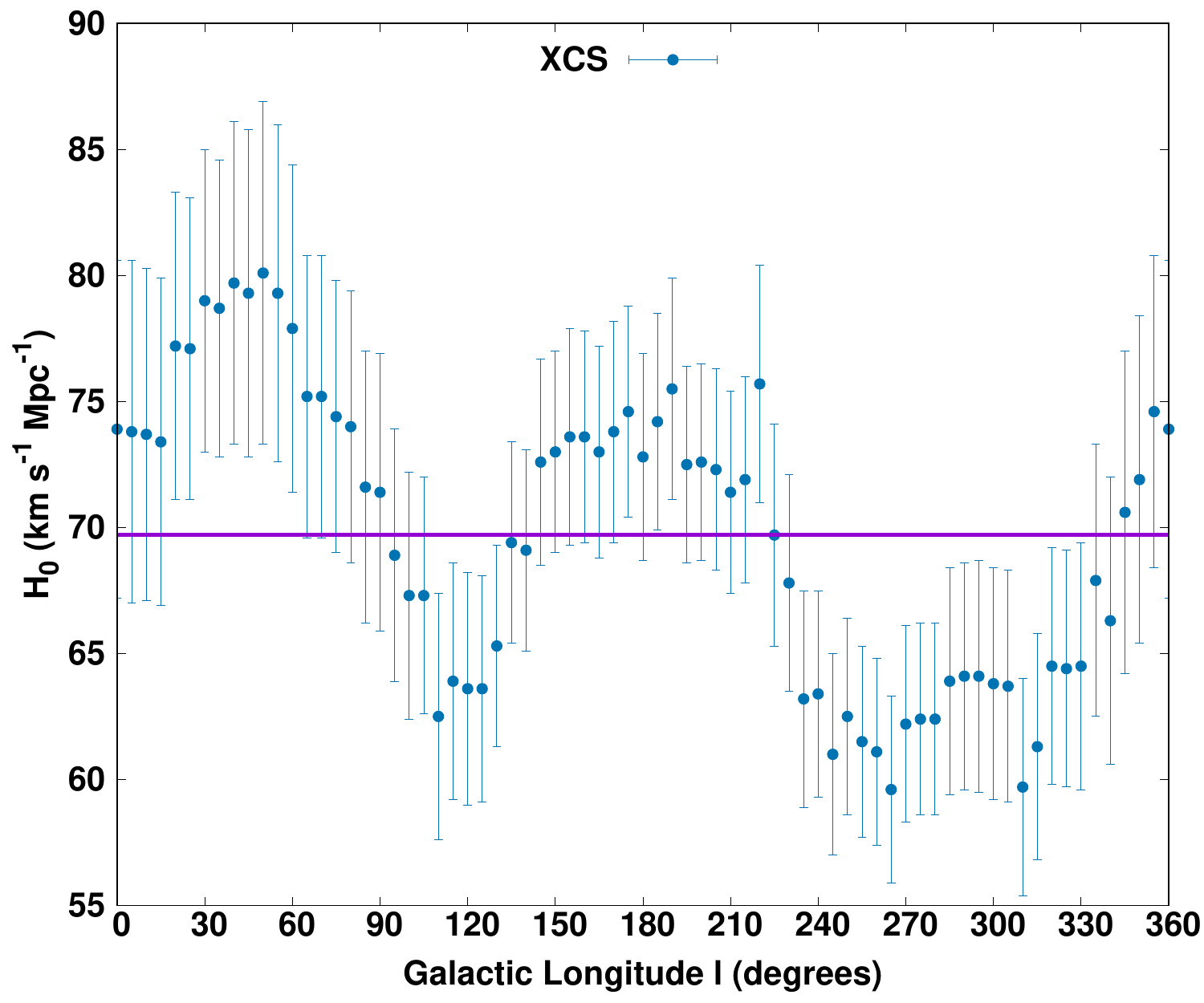}
            \includegraphics[width=0.49\textwidth, height=6cm]{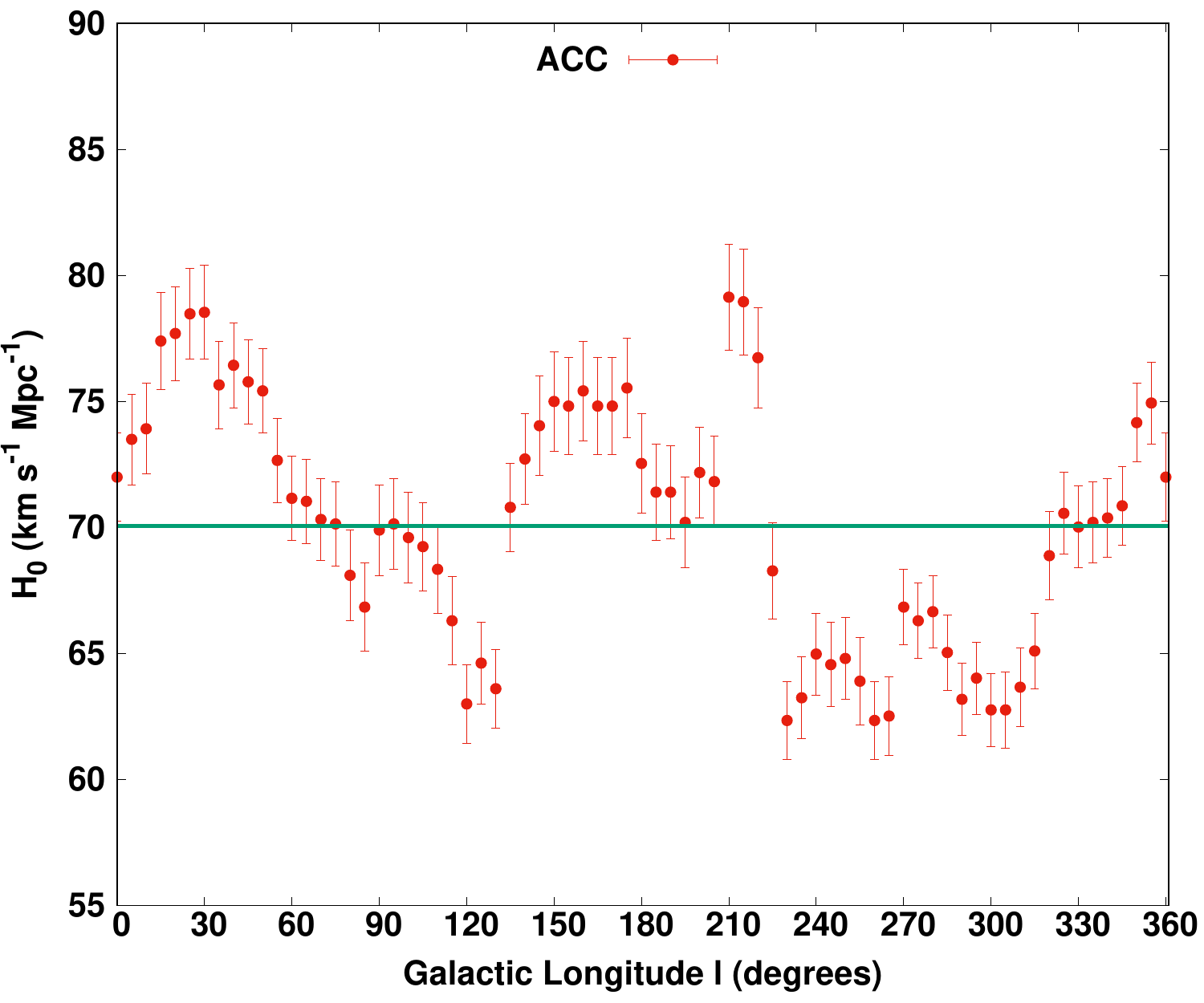}
          \caption{Best-fit value of $H_0$ for every sky region of XCS-DR1 (left) and ACC (right) with $\Delta l=90^{\circ}, \ \Delta b=180^{\circ}$ as a function of its central Galactic longitude.}
        \label{hubble-regions}
\end{figure*}

For the entire XCS-DR1 sample, we obtain $H_0=69.32\pm 2.32\ \text{km\ s}^{-1}\ \text{Mpc}^{-1}$, while the results from the sky "scanning" are shown in the left panel of Fig.  \ref{hubble-regions}. As expected, we obtain the same fluctuation pattern for $H_0$  as we did before for the normalization as well as for $\Omega_{\text{m}}$. Once again, Group A returns the largest value, namely $H_0=80.00^{+6.36}_{-5.90}\ \text{km\ s}^{-1}\ \text{Mpc}^{-1}$, which deviates by $2.76\sigma$ from the rest of the sample ($H_0=67.09\pm 2.45\ \text{km\ s}^{-1}\ \text{Mpc}^{-1}$). At the same time, for Group B we obtain $H_0=60.78^{+3.54}_{-3.36}\ \text{km\ s}^{-1}\ \text{Mpc}^{-1}$ ($2.74\sigma$) and for Group C $H_0=59.61^{+5.14}_{-4.76}\ \text{km\ s}^{-1}\ \text{Mpc}^{-1}$ ($2.41\sigma$). We see that Groups A and B have a similar statistical significance, with the difference that the Group A region has a consistent behavior regardless of the fitted parameter or the galaxy cluster sample used, unlike Group B region (although its statistical significance is $>1.5\sigma$ in all cases).

For the full ACC sample, we find $H_0=70.06\pm 0.84\ \text{km\ s}^{-1}\ \text{Mpc}^{-1}$ while the results from the sky "scanning" are shown in the right panel of Fig.  \ref{hubble-regions} . Group A appears to have the largest value (that does not depend on single outliers) with $H_0=78.30\pm 1.82\ \text{km\ s}^{-1}\ \text{Mpc}^{-1}$ ($2.36\sigma$), while Group C is the sky region with the most statistically significant small value of $H_0=63.12\pm 1.51\ \text{km\ s}^{-1}\ \text{Mpc}^{-1}$ ($2.17\sigma$). Additionally, for Group B we obtain $H_0=62.34\pm 1.52\ \text{km\ s}^{-1}\ \text{Mpc}^{-1}$ ($1.84\sigma$). Now, once more, we exclude the most extreme outlier (galaxy cluster 2A 0335+096) which, as we showed in Section \ref{sky_angles}, has the strongest effect on the entire sample due to its "problematic" temperature determination. As a result, the deviation of Group A shifts to  ($3.34\sigma$), while the one for Group C remains relatively constant.

From all the obtained results, we see that the Group A  sky region is mimicking the behavior of the minimum acceleration axis in a supposedly anisotropic universe, with a  $\gtrsim 2.5\sigma$ statistical significance in all cases. Generally, $H_0$ deviations are of slightly smaller amplitude than the ones of the normalization and $\Omega_{\text{m}}$.

\section{Discussion}

While the $L_X-T$ relation has been analyzed in several previous studies using different samples, there has not been an extensive analysis of the isotropy of this scaling relation for different sky directions. The main advantage of such an application, is the ability of directly measuring the X-ray flux, temperature, and redshift of the clusters, without the use of (cosmological) assumptions (unlike the mass determination). This classifies the $L_X-T$ relation as a very effective tool with which we can test the validity of the CP, totally independently from all the other cosmological probes used for this purpose. In addition, the almost entire independence of the two galaxy cluster samples used here gives us the opportunity to verify if a detected behavior of a sky region is consistent for every sample or if it the result of one sample's systematics.

The main finding of this project is the existence of an apparent anisotropy in the $L_X-T$ scaling relation between the so-called Group A sky region within $l\sim (-15^{\circ}, 90^{\circ})$ and the rest of the sky, and especially the sky regions within $l\sim (200^{\circ}, 310^{\circ})$ (Group B) and $l\sim (90^{\circ}, 160^{\circ})$ (Group C). The regions are slightly different for the two samples but are towards the same directions. We calculate the deviation using the Bootstrap method with 10000 random, non-identical samples every time.  The deviation of the normalization of Group A (as defined in every sample separately) is $2.65\sigma$ for ACC and $3.12\sigma$ for XCS-DR1 when compared to the rest of the sample, while it shifts to $\sim 4\sigma$ when compared to other coherent and independent sky patches. 

Another very noteworthy result is the striking similarity of the normalization trend as a function of the Galactic longitude between the two independent samples, following a pattern where high and low peaks are separated by $\Delta l\sim 90^{\circ}$. This could point to the need for an explanation that does not emanate from systematic behaviors of the samples but rather from independent factors, such as large-scale effects.

Moreover, using the Jackknife resampling method, we try to identify possible outliers that could be responsible for the deviations between subsamples. From this procedure, we see that single clusters can heavily affect the obtained result when the size of the subsample is not sufficient. 
However, for the specific groups, the tension is not significantly broken when we exclude their most "extreme" clusters. Other factors that were proven unable to explain the derived results were a redshift conversion to the CMB rest frame, a bulk flow motion of our Galaxy towards any direction and for any velocity amplitude, an exclusion of the galaxy clusters within $|b|\leq 20^{\circ}$, a possible difference in the $N_{\text{HI}}$ values of different sky regions, or different $T$ and $z$ distributions. Furthermore, using the HIFLUGCS sample, we saw that the deviation of Group A is even stronger for the brightest clusters and is not at all affected by the cool-cores of many of the members. Another possible cause that has to be tested in more detail, is the effect that nearby structures, such as superclusters, could have on the $L_X-T$ behavior of the member clusters. The simple test that we applied, indeed shows that clusters found in rich environments such as superclusters tend to have lower X-ray luminosities for the same temperature than clusters that are more isolated. However, as we pointed out, this does not seem to be the cause of the apparent deviation between Group A and the rest of the sky. The sky distribution of such structures, is roughly uniform between the two subsamples. In addition, the differences in the normalization between Group A and the rest of the sky can be also identified in the clusters that belong in superclusters (with large uncertainties of course due to the limited number of such clusters). Moreover, the dependance of the $T$ measurement on different $N_{\text{HI}}$ values, in the case when a double thermal model fitting is needed, has to be studied in the future.

When we fix the normalization and slope to the same values for the entire sky, the behavior of the fitted cosmological parameters $\Omega_{\text{m}}$ and $H_0$ is almost identical with the behavior that the normalization appeared to have before, having the same trends and deviation amplitudes. This indicates the effectiveness with which one can trace the differences in the obtained cosmological parameter values by only studying the normalization of the $L_X-T$ scaling relation. If one considers an anisotropic Hubble expansion to be the reason behind these statistically significant discrepancies between different sky regions, then Group A seems to have the minimum expansion rate, in contrast with Groups B (mainly) and C that appear to have the opposite behavior.

On the other hand, it should be pointed out here that galaxy clusters are driven by complex physical processes while the $L_X-T$ relation generally suffers from a large intrinsic scatter, and therefore other physical reasons for these apparent deviations must be examined. 
A future comparison with a large number of data from cosmological simulations would give us a better idea of how likely these deviations are to occur within a cosmologically isotropic frame.

Additionally, future all-sky samples such as eeHIFLUGCS \citep{eeHIF} and the one that will be constructed with the eROSITA telescope \citep{merloni,pillepich,borm,erosita16}, would be extremely useful to further investigate this phenomenon in depth. Finally, since the new method that we introduce in this paper is used for the first time, more work is needed to properly assess the significance of the obtained results.

\section{Conclusions}

In this study, we suggest a new method which takes advantage of the $L_X-T$ scaling relation of galaxy clusters in order to investigate the validity of the Cosmological Principle. Based on the fact that cosmological parameters enter the relation only through the luminosity distance and the redshift evolution of the $L_X-T$ relation, any deviations in the normalization of the latter will be strongly correlated with deviations in the cosmological parameters.

For this purpose, we use two full-sky galaxy cluster samples, ACC and XCS-DR1, which share less than $1\%$ of their clusters. Remarkably, we find that the two samples have their higher and lower normalization values at similar Galactic longitudes, showing the same fluctuation tendency. The sky region that stands out the most in both samples is the one within $l\sim (-15^{\circ},90^{\circ})$ (Group A) that appears to have a statistically significant deviation from the rest of the sample ($\sim 3\sigma$). This behavior is consistent in both ACC and XCS-DR1 samples. More specifically, if one compares the regions with the lowest $a$ values, Group A seems to have a $\sim 4\sigma$ deviation.

Several possible reasons for these inconsistencies were considered. Firstly, we demonstrated that single clusters that act as outliers are not the reason for the peculiar behavior of Group A region. Excluding clusters in low Galactic latitudes that could suffer from an improper $L_X$ derivation due to absorption did not alleviate the tension between the best-fit results of the independent subsamples. Furthermore, using the HIFLUGCS sample whose vast majority of clusters are contained in ACC, we saw that cool-core clusters are not responsible for the differences between Group A and the rest of the sky. A specific structure in the hydrogen column densities of different sky directions and its effects on the obtained $L_X$ values does not seem to be the case either, although the same remains to be verified for $T$ measurements also. A redshift conversion from heliocentric with respect to the CMB frame as well as redshift conversions to account for possible bulk flows of the clusters, were also applied to both samples, without significantly changing the results. Moreover, Group A has similar $T,\ z$ and $\sigma_{\log{Lx,\ T}}$ distributions to the rest of the sky and to the lowest-$a$ regions, for both ACC and XCS-DR1. Another reason that was tested is the effect of the environment on the $L_X-T$ behavior of galaxy clusters. While we found that clusters that appear to belong to superclusters tend to have lower $L_X$ values compared to field clusters, this cannot explain the observed deviations.

The reason behind this apparent mild anisotropy in the normalization value of the $L_X-T$ relation remains to be identified. This could also be expressed as a difference in the luminosity distance, for example, cosmological parameters of different sky regions. The same fluctuation trend as for the normalization has been obtained for $\Omega_{\text{m}}$ and $H_0$  for both samples, while the statistical significance of Group A is similar in all cases ($\sim 3\sigma$). Of course, in order for such an extreme scenario to be considered, the statistical significance of the results must be higher and all the possible physical reasons for such a deviation must be examined and eventually rejected. Comparisons with isotropic cosmological simulations of $>10^6$ galaxy clusters will also be made in future work. Finally, next-generation samples that will cover the full sky, such as eeHIFLUGCS and the eROSITA all-sky survey sample, containing large amounts of data with exceptional accuracy, will allow us to further investigate these interesting behaviors.

\begin{acknowledgements}  
We thank the anonymous referee for their useful and constructive comments which helped us improve the quality of the paper. Moreover, we thank Florian Pacaud for providing us with the supercluster detection results using his code, as well as for his very helpful advice. KM is a member of the Bonn-Cologne Graduate School for Physics and Astronomy(BCGS), and thanks for its support. THR acknowledges support by the German Research Association (DFG) through grant RE 1462/6 and through the Transregional Collaborative Research Centre TRR33 The Dark Universe (project B18) as well as by the German Aerospace Agency (DLR) with funds from the Ministry of Economy and Technology (BMWi) through grant 50 OR 1608. 
This research has made use of the NASA/IPAC Extragalactic Database (NED)
which is operated by the Jet Propulsion Laboratory, California Institute of Technology, under contract with the National Aeronautics and Space Administration.
\end{acknowledgements}


\end{document}